%% file: main.tex
\documentclass[10pt,journal,compsoc]{IEEEtran}
\usepackage[ruled,linesnumbered,resetcount]{algorithm2e}
\usepackage{tcolorbox}
\usepackage{algpseudocode}
\usepackage{amssymb}
\usepackage{amsthm}
\usepackage{amsmath}
\usepackage{caption}
\usepackage{subcaption}
\newtheorem{observation}{Observation}
\newtheorem{definition}{Definition}
\usepackage{multicol}
\usepackage{balance}
\usepackage{multirow}
\usepackage{url}
\usepackage{graphicx}
\newcommand{\modify}[1]{
  {\color{black} #1}}

\newcommand{\tool}{\textsc{Invalidator}}
\newcommand{\ods}{\textsc{ODS}}
\newcommand{\rgt}{\textsc{RGT}}

\ifCLASSINFOpdf
\else
\fi
\begin{document}
%
\title{Invalidator: Automated Patch Correctness Assessment via Semantic and Syntactic Reasoning}

\author{\IEEEauthorblockN{Thanh Le-Cong,
Duc-Minh Luong, Xuan Bach D. Le,
David Lo, 
\\ Nhat-Hoa Tran, Bui Quang-Huy and
Quyet-Thang Huynh}
\IEEEcompsocitemizethanks{
\IEEEcompsocthanksitem Thanh Le-Cong, Xuan-Bach D. Le are  with School of Computing and Information Systems, The University of Melbourne
\\
E-mail: congthanh.le@student.unimelb.edu.au, bach.le@unimelb.edu.au, 
\IEEEcompsocthanksitem Duc-Minh Luong, Quang-Huy Bui, Nhat-Hoa Tran and Quyet-Thang Huynh are with School of Information and Communication Technology, Hanoi University of Science and Technology \\
E-mail: \{minh.ld176821, huy.bq20202517M\} @sis.hust.edu.vn, \{hoatnt, thanghq\}@soict.hust.edu.vn
\IEEEcompsocthanksitem David Lo is with School of Computing and Information Systems, Singapore Management University\\
E-mail:  davidlo@smu.edu.sg
\IEEEcompsocthanksitem Quyet-Thang Huynh is the corresponding author.

}

}

%



\IEEEtitleabstractindextext{%
\begin{abstract}
Automated program repair (APR) has been gaining ground recently. However, a significant challenge that still remains is patch overfitting, in which APR-generated patches plausibly pass the validation test suite but fail to generalize. A common practice to assess the correctness of APR-generated patches is to judge whether they are equivalent to ground-truth, i.e., developer-written patches, by either generating additional test cases or employing human manual inspections. The former often requires the generation of at least one test witnessing the behavioral differences between the APR-patched and developer-patched programs. Searching for the witnessing test, however, can be difficult as the search space can be enormous. Meanwhile, the latter is prone to human biases and requires repetitive and expensive manual effort. 

In this paper, we propose a novel technique, namely \tool{}, to automatically assess the correctness of APR-generated patches via semantic and syntactic reasoning. \tool{} reasons about program semantic via program invariants while it also captures program syntax via language semantic learned from large code corpus using the pre-trained language model. Given a buggy program and the developer-patched program, \tool{} infers likely invariants on both programs. Then, \tool{} determines that a APR-generated patch overfits if: (1) it violates correct specifications or (2) maintains errors behaviors of the original buggy program. In case our approach fails to determine an overfitting patch based on invariants, \tool{} utilizes a trained model from labeled patches to assess patch correctness based on program syntax. The benefit of \tool{} is three-fold. First, \tool{} is able to leverage both semantic and syntactic reasoning to enhance its discriminant capability. Second, \tool{} does not require new test cases to be generated but instead only relies on the current test suite and uses invariant inference to generalize the behaviors of a program. Third, \tool{} is fully automated. We have conducted our experiments on a dataset of 885 patches generated on real-world programs in Defects4J. Experiment results show that \tool{} correctly classified 79\% overfitting patches, accounting for 23\% more overfitting patches being detected by the best baseline. \tool{} also substantially outperforms the best baselines by 14\% and 19\% in terms of Accuracy and F-Measure, respectively.
\end{abstract}

}

\maketitle

\IEEEdisplaynontitleabstractindextext

%
\IEEEpeerreviewmaketitle

\section{Introduction}\label{sec:intro}
\input{contents/introduction}

\section{Background}\label{sec:background}
\input{contents/background}
\section{Motivation}\label{sec:motivation}
\input{contents/example}
\section{Methodology}\label{sec:model}
\input{contents/methodology}

\section{Empirical Evaluation}\label{sec:eval}
\input{contents/evaluations}
\section{Discussion} \label{sec:threats}
\input{contents/thread}
\section{Related Work}\label{sec:relatedworks}
\input{contents/relatedworks}
\section{Conclusion and Future Work} \label{sec:conclusion}
\input{contents/conclusion}



\ifCLASSOPTIONcaptionsoff
  \newpage
\fi

\balance
\bibliographystyle{IEEEtran}
\bibliography{main.bib}

%








\end{document}

%% file: contents/introduction.tex
Automated program repair (APR) is a promising approach to alleviate the onerous burden on developers to manually fix bugs. Over the years, various APR techniques have been proposed~\cite{le2011genprog, kim2013automatic, qi2015analysis, le2016history, xuan2016nopol, long2016automatic, mechtaev2016angelix, xiong2017precise, le2017s3, mahajan2018automated, liu2018lsrepair, liu2019tbar, chen2019sequencer, le2021usability,  ye2022neural, xia2022less, ye2022selfapr}, with several breakthroughs that inspired potential practical adoption of APR. Notably, Facebook has recently deployed SapFix~\cite{marginean2019sapfix}, the first-ever industrial-scale automatic bug-fixing system, for suggesting fixes to developers in real-world products. Despite these recent successes, APR still suffers from a major challenge, namely \textit{test overfitting}~\cite{smith2015cure,le2018overfitting, nilizadeh2021exploring}, in which a generated patch may pass all test cases but still fails to generalize to the intended behaviors of the program. According to Qi et al.~\cite{qi2015analysis} 98\% of the plausible patches generated by GenProg~\cite{le2011genprog} are overfitting. 

\par Detecting overfitting patches is one key challenge that is important not only to ensure fair comparisons between APR techniques but also to enable the practical adoption of APR by developers. Often, one APR technique claims to be better than others only solely in terms of the number of bugs for which it can generate ``correct'' patches. Furthermore, recent research suggested that low-quality patches may negatively affect developers' performance~\cite{tao2014automatically}. A fundamental question then arises,
\begin{center}
    "\textit{How can we determine whether a patch is correct?}"
\end{center}
Unfortunately, even with the availability of the ground truth (developer-patched) program, it is difficult to determine whether an APR-patched program is correct. That is because, unless the APR-patched program and the ground truth program are exactly syntactically the same, determining whether two programs are semantically equivalent is indeed an undecidable problem~\cite{kozen1977rice}. 

Recent approaches in automated program repair (APR) have explored various techniques to assess the correctness of APR-generated patches in comparison to developer-written patches as ground truth. These approaches include the use of test-suite augmentation and human manual inspections. While being human manual inspections are effective~\cite{le2019reliability}, they are prone to human biases, are expensive, and require manual, repetitive tasks. On the other hand, test-suite augmentation approaches, such as those proposed by Xin et al.~\cite{xin2017identifying} and Xiong et al.~\cite{xiong2017precise}, are fully automated, but they require the generation of at least one test case to observe behavioral differences between the APR-patched program and the ground truth program. However, a recent study~\cite{le2019reliability} has shown that test-suite augmentation approaches are often ineffective as the search space for bug-witnessing test cases can be large. Even state-of-the-art test case generation techniques, such as Randoop~\cite{pacheco2007randoop} and \textsc{DiffTGen}~\cite{xin2017identifying}, can only identify 22\% of the overfitting patches generated by APR tools~\cite{le2019reliability}. Furthermore, Xin et al.~\cite{xin2017identifying} reported that in many cases, the state-of-the-art test generator \textsc{EvoSuite}~\cite{fraser2011evosuite} failed to generate any test methods that exercise the code changes introduced in the generated patches, and thus failed to identify behavioral differences between the patches and the ground truth.

\par In this paper, we introduce a novel technique called \tool{} that combines semantic and syntactic reasoning to automatically assess the correctness of patches generated by APR techniques. \tool{} leverages program invariants to reason about program semantics and pre-trained language models to capture program syntax by learning language semantics from a large code corpus. Similar to other automated patch correctness assessment (APAC) techniques, \tool{} utilizes behavioral discrepancies between the APR-patched and ground truth programs to determine the patch's correctness. However, conceptually, \tool{} is different from the strategy employed by existing APAC techniques such as \textsc{DiffTGen}~\cite{xin2017identifying}, \textsc{PatchSim}~\cite{xiong2018identifying}, or \textsc{Randoop}~\cite{pacheco2007randoop, le2019reliability}. These techniques generate new tests to augment the current test suite, in which each test generates one execution. As a result, the chance to hit an execution that reveals a behavioral difference between the APR-patched and ground truth programs is approximately linearly proportional to the number of tests generated. In contrast, \tool{} only uses the current test suite and infers program invariants that naturally generalize beyond the test suite. The generalization of program invariants allows \tool{} to effectively and semantically reason about program correctness. Additionally, \tool{} further augments program semantic reasoning by incorporating syntactic reasoning to enhance its effectiveness. We describe the details of the semantic and syntactic reasoning in \tool{} below.

Given an APR-generated patch, the original buggy program, and its correct (ground truth) version, \tool{} works in two main phases. 

\noindent \textcircled{1} \textbf{Semantic-based Classifier.} The semantic-based classifier is built based on two high-level intuitions. 
First, program invariants that are maintained in both the buggy and correct (ground truth) versions of a program can serve as the \textit{correct} specifications of the program. 
Second, program invariants that only exist in the buggy program but do not exist in the correct version may represent the \textit{error} specifications of the program. \tool{} determines that a machine-generated patch overfits if the machine-patched program: (1) violates correct specifications or (2) maintains error specifications. 
Particularly, \tool{} first automatically infers likely invariants of each program based on its original test suite by using \textsc{Daikon}~\cite{ernst2007daikon}, a well-known invariant inference tool. \tool{} then constructs the set of {\em correct and error specifications}, which serve as approximate specifications for the program under test. 
Based on the inferred specifications, \tool{} determines that a patch is overfitting if invariants inferred from the machine-patched program either violate the correct specifications or maintain error specifications.

\vspace{2mm}

\noindent \textcircled{2} \textbf{Syntactic-based Classifier.} In case the invariant-based specification inference fails to determine an overfitting patch, \tool{} further the overfitting patches via language semantic differences between the machine-generated patch and its buggy and correct version.
Specifically, \tool{} employs a pre-trained language model, namely \textsc{CodeBert} to extract source syntactic features from the source code of each program. 
\tool{} then measures the differences by a set of comparison functions, e.g., subtraction or similarity. Finally, \tool{} uses a trained model from labeled data to estimate the likelihood of the machine-generated patch being overfitting based on the syntactic proximity.

\par We conducted our experiments on a dataset of 885 patches which include 508 overfitting patches and 377 correct ones generated for large real-world programs in the Defects4J dataset. To investigate the effectiveness of our approach, we compared \tool{} against the state-of-the-art APAC techniques, consisting of RGT~\cite{ye2021automated}, \ods{}~\cite{ye2019automated}, BERT+LR~\cite{tian2020evaluating}, \textsc{PatchSim}~\cite{xiong2018identifying}, \textsc{DiffTGen}~\cite{xin2017identifying}, \textsc{Anti-patterns}~\cite{tan2016anti}, \textsc{Daikon}~\cite{yang2020exploring}. Experiment results showed that \tool{} \modify{correctly classified} 79\% of overfitting patches, accounting for 23\% more overfitting patches being detected as compared to the best baseline. \tool{} also remarkably outperforms the best baselines by 14\% (0.81 vs. 0.68) and 19\% (0.87 vs. 0.76) in terms of \textit{Accuracy} and \textit{F1-score}, respectively.

\par In summary, we made the following contributions:
\begin{itemize}
        \item We introduced \tool{}, a novel technique that uses both semantic reasoning (via program invariants) and syntactic reasoning (via source code features) to automatically assess APR-generated patches. Our empirical evaluation demonstrated that our approach effectively detects 79\% overfitting patches with a precision of 97\%.
        
        \item We introduced two overfitting rules that rely on program invariants to assess APR-generated patches. Our empirical evaluation demonstrated that these rules can effectively identify 51\% of overfitting patches with a precision of 97\%.
        
        \item We proposed using syntactic reasoning from the program source code to augment semantic reasoning from the two aforementioned overfitting rules. Our empirical evaluation showed that syntactic reasoning can boost the performance of our approach by 35\% and 30\% in terms of Accuracy and F1-score, respectively.
        
        \item We conducted experiments on 885 machine-generated patches for the Defects4J benchmark. The experiment results showed that the unique combination of syntactic and semantic reasoning empowers \tool{} to achieve substantial improvements (i.e., 19\% and 14\% for Accuracy and F1-score, respectively) over state-of-the-art baselines. 
        
\end{itemize}

\par The rest of this paper is organized as follows: Section~\ref{sec:background} provides the background information on the overfitting problem, APAC, and program invariants. Section~\ref{sec:motivation} presents a motivating example for our approach, followed by Section~\ref{sec:model} which describes our approach in detail. Section~\ref{sec:eval} presents our experimental setup and results. In addition, Section~\ref{sec:threats} discusses the efficiency, potential applications, and threats to validity of our approach. Section~\ref{sec:relatedworks} provides an overview of related work in this area. Finally, in Section~\ref{sec:conclusion}, we conclude our paper and discuss future work.

%% file: contents/background.tex
This section presents an outline of recent automated program repair (APR) techniques and the overfitting problem in APR and discusses techniques for assessing the correctness of APR-generated patches. We subsequently discuss program invariants and dynamic invariant inference. 

\subsection{Automated Program Repair}\label{sec:auto_patch_assess}
\textbf{Program Repair.} Given a buggy program and a set of test cases in which there exists at least one failing test, the overall goal of automated program repair (APR) techniques is to generate a patch that passes all the test cases while not introducing new bugs. Generally, APR techniques can be categorized into two main families, including search-based repair and semantic-based repair. Search-based techniques often use meta-heuristic algorithms, e.g. genetic programming~\cite{le2011genprog}, random search~\cite{qi2014strength}, or learning algorithms such as data mining and machine learning, e.g.,~\cite{le2016history},~\cite{koyuncu2020fixminer},~\cite{chen2019sequencer}, and~\cite{xiong2017precise}, to apply mutations and evolve the buggy program until they find a patch passing the test suite. Semantics-based repair techniques, e.g. \textsc{Angelix}~\cite{mechtaev2016angelix}, \textsc{S3}~\cite{le2017s3}, \textsc{JFix}~\cite{le2017jfix}, use semantic analysis, e.g. symbolic execution, and program synthesis to construct patches that satisfy certain semantic constraints. We will elaborate in detail on these techniques in the related work section (Section~\ref{sec:relatedworks}).
\vspace{0.2cm}

\noindent\textbf{Overfitting.} One primary challenge in automated program repair (APR) is that APR-generated patches can be tests-adequate but may not generalize. This phenomenon, known as the 'test overfitting' problem, refers to situations where APR-generated patches successfully pass all test cases but are not semantically correct, as demonstrated in prior work~\cite{qi2015analysis, smith2015cure, le2018overfitting, nilizadeh2021exploring}. Early APR techniques utilized an existing test suite as an oracle to evaluate the correctness of generated patches~\cite{weimer2009automatically, le2011genprog}. Specifically, a patch is considered correct if it successfully passes all test cases and incorrect otherwise~\cite{weimer2009automatically,le2011genprog}. However, recent studies~\cite{qi2015analysis,smith2015cure} showed that this assessment method is insufficient to ensure the correctness of generated patches, as the test suite used for evaluation is often incomplete. Through manual analysis, Qi et. al.~\cite{qi2015analysis} have shown that the majority of patches generated by search-based APR techniques, such as \textsc{GenProg}~\cite{le2011genprog}, \textsc{AE}~\cite{weimer2013leveraging}, and \textsc{RSRepair}~\cite{qi2014strength} exhibit overfitting. Similarly, through automated evaluation, Le et al.~\cite{le2018overfitting} also have reported that semantic-based repair techniques, such as \textsc{Angelix}~\cite{mechtaev2016angelix} are no exception to the overfitting issue.
\vspace{2mm}

\noindent\textbf{Automated Patch Correctness Assessment.} Recently, researchers have often adopted one of two approaches for assessing the correctness of program repairs: (1) manual annotation, where the authors of repair techniques manually judge the correctness of APR-generated patches by their own and competing approaches, or (2) automated assessment, where an independent test suite is used to automatically evaluate patch correctness. However, Le et al.~\cite{le2019reliability} showed that while a manual annotation is more effective, it is also more expensive. In contrast, an automated assessment does not require a manual effort but is less effective~\cite{le2019reliability}. Recent research efforts have been devoted to automated patch correctness assessment (APCA)~\cite{xin2017identifying, xiong2018identifying, le2019reliability}. Existing APCA techniques usually assume that ground truth patches are available for comparison~\cite{xin2017identifying, le2019reliability, yu2019alleviating}. For example, Xin et al.~\cite{xin2017identifying}, and Le et al.~\cite{le2019reliability} generate new test cases based on the program (i.e., ground truth) to identify overfitting patches. Our proposed technique also falls into this category, where we assume the availability of ground truth patches. However, unlike existing test-based approaches, our technique relies on program invariants to judge the correctness of APR-generated patches.

\subsection{Program invariants\label{sec:invariant}}

Program invariants \textit{(invariants for short)} is a term referring to \textit{properties that hold at a certain program point or points, which might be found in an $assert$ statement, or a formal specification}~\cite{ernst2007daikon}. For example, a program invariant can be $ x>= abs \left( y \right)$ or $size\left(A\right) == size\left(B\right)$. Among several of their usages, program invariants can be used to detect modifications that violate the original properties of a program. 

True invariants, however, are usually difficult to obtain in real-world projects, and thus researchers often resort to properties known as likely invariants, which \textit{hold for some executions, but perhaps not all}~\cite{b2016learning,ding2019leveraging}. Likely invariants can be automatically inferred from execution traces by dynamic invariant inference techniques which generalize from execution traces using invariant templates. Previous studies have demonstrated the effectiveness of likely invariants in various tasks including complexity analysis~\cite{nguyen2017counterexample, ishimwe2021dynaplex}, termination analysis~\cite{le2020dynamite}, bug localization~\cite{le2016issta} and neural network analysis~\cite{gopinath2019property, nguyen2022gnninfer}. 

In this paper, we use Daikon~\cite{ernst2007daikon} - a popular tool for mining likely invariants, as our dynamic invariant inference technique. Daikon observes the execution traces of programs and matches them against a set of templates to infer likely invariants that hold on all or most of the executions. From a large set of 311 templates (c.f. Details in Daikon Manual Documentation \footnote{\url{http://plse.cs.washington.edu/daikon/download/doc/daikon/Daikon-output.html#Invariant-list}}), Daikon can detect a wide variety of invariants that generalize well beyond the test suite used to produce the execution traces~\cite{sagdeo2013using}. 

%% file: contents/example.tex
\begin{figure}[!htb]
\centering
\begin{subfigure}[H]{\columnwidth}
\includegraphics[width=3.2in]{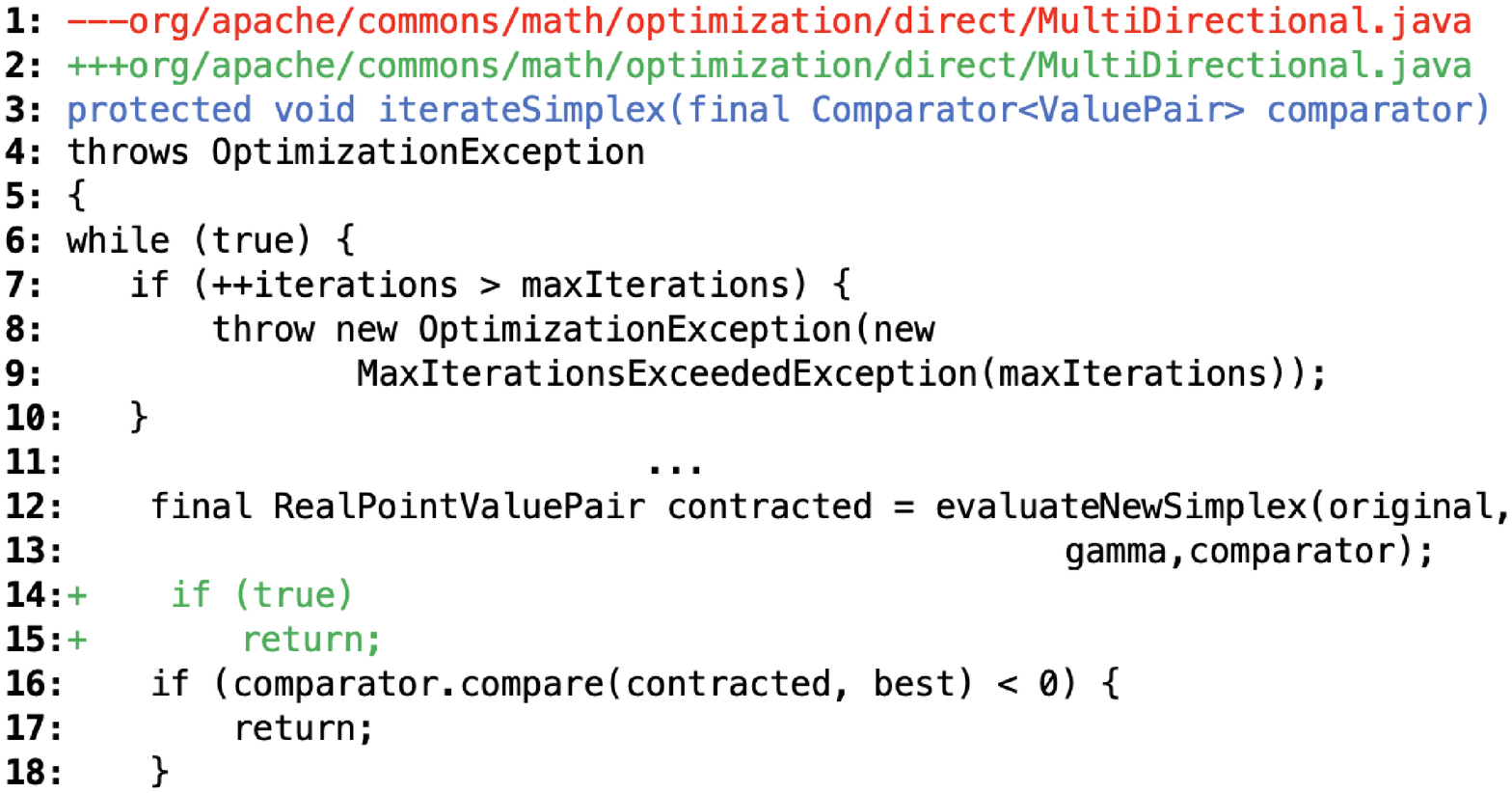}
\hfill
\caption{An overfitting patch generated by Kali~\cite{qi2014strength} \label{fig:motivation_machine}}
\end{subfigure}

\vspace{3mm}

\begin{subfigure}[H]{\columnwidth}
\includegraphics[width=3.2in]{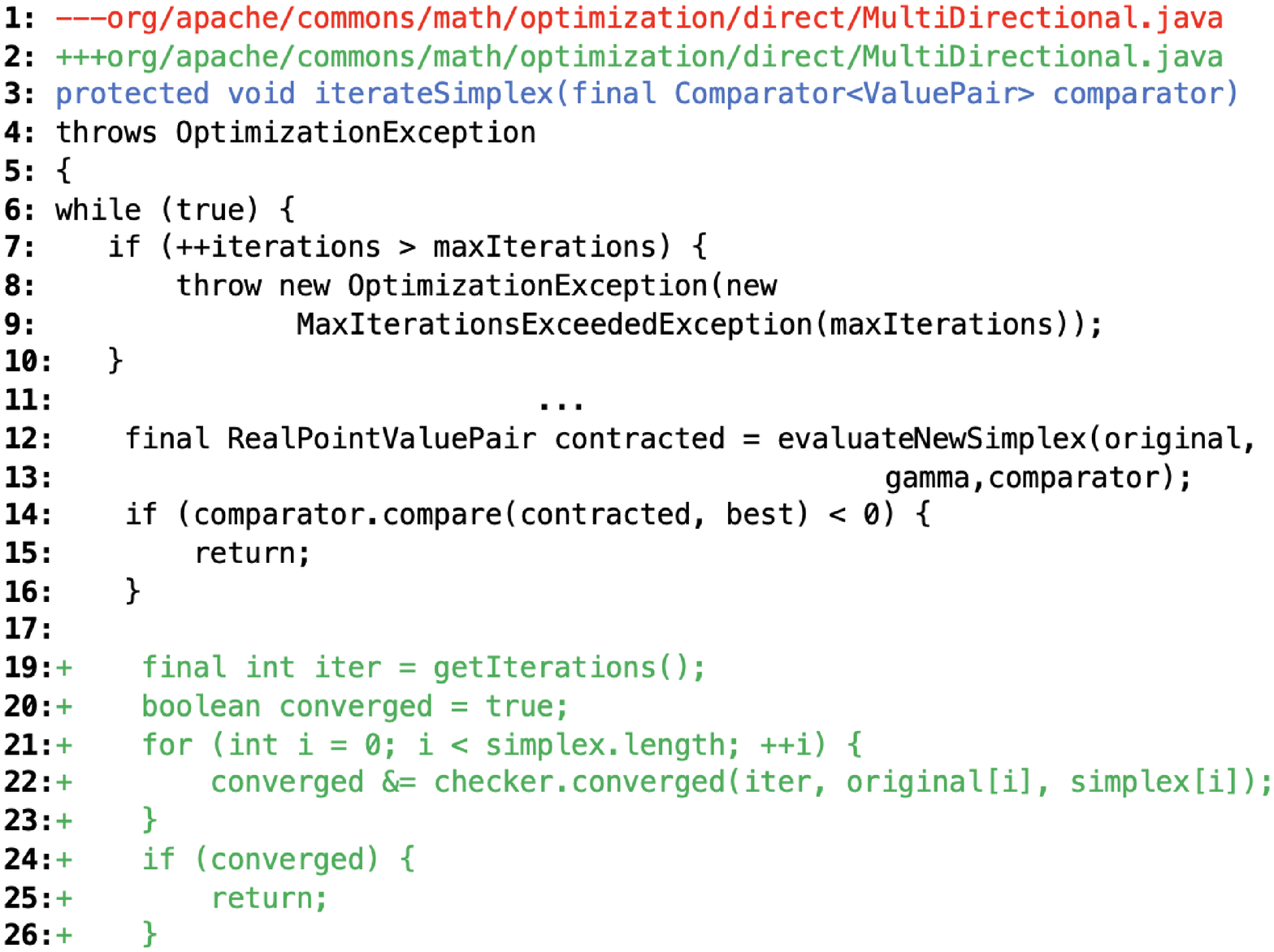}
\caption{The correct patch written by human developers \label{fig:motivation_human}}
\end{subfigure}
\caption{An overfitting patch generated by Kali and the ground truth human-written patch for Math-84}
\label{fig:motivation}
\end{figure}

Let us now use an example to motivate our approach of using program invariants to determine patch correctness. The bug example in Figure~\ref{fig:motivation} shows an APR-generated patch (Figure~\ref{fig:motivation_machine}) and the ground truth developer-written patch (Figure~\ref{fig:motivation}b). 

\begin{figure*}[h]
    \centering
    \includegraphics[width=6.4in]{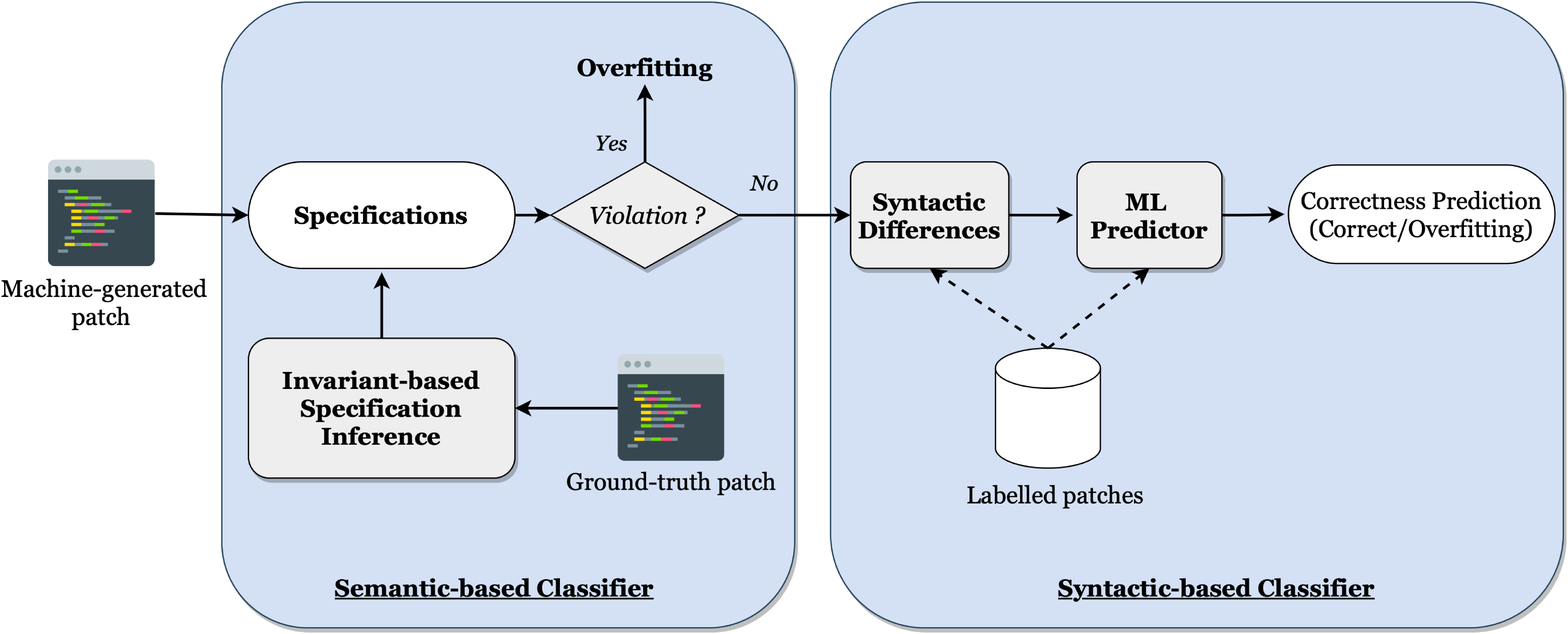}
    \caption{The workflow of \tool{}}
    \label{fig:overview}
\end{figure*}

In this example, the \texttt{iterateSimplex()} method contains a bug that causes an endless loop (line 6 in Figure~\ref{fig:motivation_human}) when computing the next simplex of the multi-directional optimizer. To fix this issue, a developer-written patch was added (line 5 and lines 19-26 in Figure~\ref{fig:motivation_human}) that stops the loop once the algorithm converges. In contrast, the plausible patch generated by \textit{Kali}~\cite{qi2014strength} (Figure~\ref{fig:motivation_machine}) inserts an early \texttt{return} at lines 14-15, causing the failing test case to plausibly pass. While the APR-patched program avoids the endless loop, it ignores the main algorithm, which requires many iterations to converge. Unfortunately, current state-of-the-art test-based automatic program repair techniques, such as \rgt{}~\cite{ye2021automated}, \textsc{DiffTGen}~\cite{xin2017identifying}, and \textsc{Randoop}~\cite{pacheco2007randoop}, have difficulty identifying behavioral differences between the APR-patched program and the ground truth~\cite{le2018overfitting} due to the large search space of bug-witnessing test cases.
\vspace{0.2cm}

\noindent \textbf{Invariants come into play.} Let us now look at how program invariants can show that the APR-patched program is overfitting. The intended behavior of the program is for the algorithm to terminate after several iterations once it converges, regardless of the input. For the \texttt{iterateSimplex} method, an invariant \texttt{iterations > orig(iterations)} is inferred from both the buggy and correct versions of the program. This invariant indicates that the value of the \texttt{iterations} variable before calling the \texttt{iterateSimplex} method (denoted as \texttt{orig(...)}) is smaller than the value after execution. This variable measures the number of iterations executed by the algorithm and reflects the fact that the while loop (Line 4 in Figure~\ref{fig:motivation_human}) should execute until the multi-directional optimizer converges, which may require many iterations.

However, in the APR-patched program, the while-loop always terminates after the first iteration due to the code snippet \texttt{if (true) { return;}} (line 14-15 in Figure~\ref{fig:motivation_machine}). Consequently, an invariant \texttt{iterations - orig(iterations) - 1 == 0} is obtained for the APR-patched program. This invariant indicates that the value of the \texttt{iterations} variable is always incremented by one after executing the \texttt{iterateSimplex} method. This behavior shows that the APR-patched program is overfitting and violates the intended behavior of the program, which requires varying numbers of iterations to converge under different inputs.

Our tool, \tool{}, detects this overfitting behavior of the APR-patched program by comparing the invariant generated from the APR-patched program with that of the buggy and ground truth programs. Specifically, \tool{} identifies that the APR-patched program maintains an invariant that never holds in the buggy and ground truth programs, indicating a behavioral divergence that could lead to errors.

%% file: contents/methodology.tex
Figure~\ref{fig:overview} illustrates the workflow of \tool{}. First, an APR-patched program is validated using a semantic-based classifier. During this phase, \tool{} assesses the correctness of the patch based on {\em correct} and {\em error specifications}, which are inferred by analyzing the differences in behavior between the buggy program and its correct version. These specifications are captured using automatically-inferred program invariants. If the invariants inferred from the APR-patched program violate the correct specification or maintain the error specifications, the APR-patched program is considered {\em overfitting}. If the inferred specifications fail to identify an overfitting patch, \tool{} uses a learning-based model that leverages syntactic reasoning to estimate the probability that the APR-patched program is overfitting. We provide further details about \tool{} below.

\subsection{Semantic-based Patch Classifier} \label{sec:hardclassifier}

Figure~\ref{fig:hard} illustrates how \tool{} employs a semantic-based patch classifier to identify {\em overfitting patches}. First, \tool{} constructs approximate specifications of the program under test by using a dynamic invariant inference tool, called \textsc{Daikon}~\cite{ernst2007daikon} (described in Section~\ref{sec:ivariantmining}). Then, based on the inferred specifications, \tool{} automatically classifies whether an APR-patched program is overfitting (explained in Section~\ref{sec:classfication}). We provide detailed explanations of each phase of \tool{} below.

\subsubsection{Invariant-based Specification Inference} \label{sec:ivariantmining}
The goal of invariant inference is to derive specifications that can determine the correct and error behaviors of a program. \tool{} employs invariants to approximate these specifications based on two key observations that enable the detection of behavioral differences, as outlined below:

\begin{observation}
Program invariants that are maintained in both the buggy and correct (ground truth) versions of a program can serve as the \textit{correct} specification for the original program.
\end{observation}

\begin{observation}
Program invariants that exist only in the buggy program, but do not hold in the correct version, may represent the \textit{error} specification of the buggy program.
\end{observation}
Based on the correct and error specification, \tool{} can heuristically assess the patch correctness. Below, we formally define the correct and error behaviors, explain how we construct them via invariant inference, and how they can be used to determine overfitting patches effectively.

The correct behaviors of a program are defined based on invariants in Definition~\ref{lbl:def_correct}. Correct behaviors reflect common behaviors in both the original buggy program and the correct (ground truth) program in which the bug is fixed by developers. We use a set of invariants, denoted as $\mathcal{C}$, which commonly appears in both the buggy version and the correct version of a program, to approximate the correct behaviors of the program. The use of $\mathcal{C}$ reduces the false positive rate (as we shall see in Section~\ref{sec:eval}).

Error behaviors are defined in Definition~\ref{lbl:def_error}, and capture the behavioral divergence of the buggy program from the correct program. The behavioral difference is reflected by a set of program invariants $\mathcal{E}$ that hold in the buggy program but do not hold in the correct version of the program.

\begin{figure}[bt]
    \centering
    \includegraphics[width=\columnwidth]{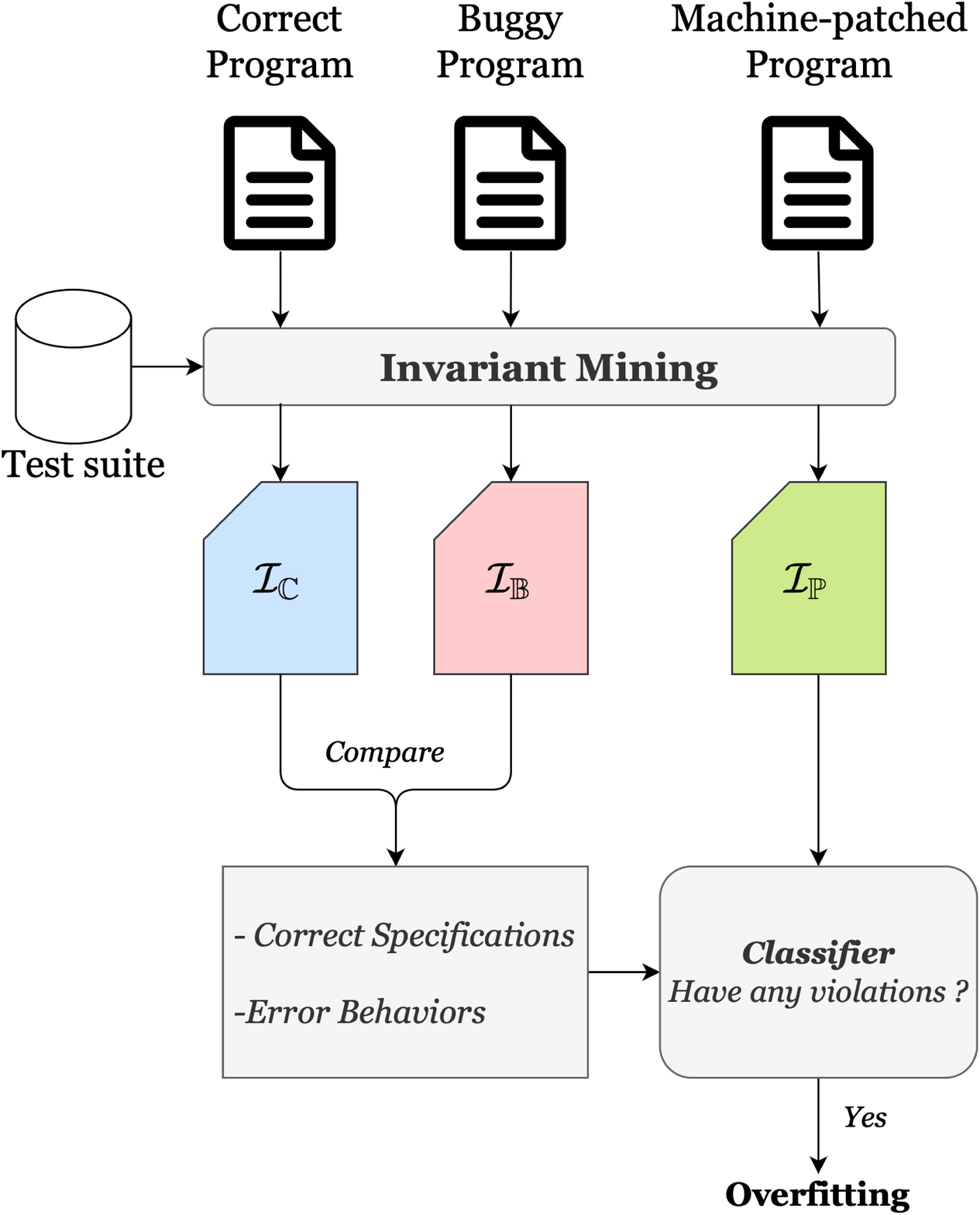}
    \caption{APAC via invariant-based specification inference. $\mathcal{I}_{\mathbb{C}}$, $\mathcal{I}_{\mathbb{B}}$ and $\mathcal{I}_{\mathbb{P}}$ are sets of invariants inferred from correct program, buggy program and APR-patched program, respectively.}
    \label{fig:hard}
\end{figure}

\begin{definition}\label{lbl:def_correct}
(Correct specification) Consider a buggy program $\mathbb{B}$ and its correct/ground truth version $\mathbb{G}$. The correct specification of $\mathbb{G}$ is approximated by a set of invariants $\mathcal{C}$ such that $\mathcal{C} \models \mathbb{B}$ and $\mathcal{C} \models \mathbb{G}$, where X $\models$ Y denotes a semantic logical consequence relation in which all properties satisfying X also satisfy Y.
\end{definition}
\begin{definition}\label{lbl:def_error} (Error specification) Consider a buggy program $\mathbb{B}$ and its correct/ground truth version $\mathbb{G}$. The error specification of $\mathbb{B}$ is approximated by a set of invariants $\mathcal{E}$ such that $\mathcal{E} \models \mathbb{B}$ and $\mathcal{E} \not\models \mathbb{G}$, where X $\models$ Y denotes a semantic logical consequence relation in which all properties satisfying X also satisfy Y.
\end{definition}

Let us now explain how we construct the specifications that approximate the correct and error behaviors of a program as depicted in Figure~\ref{fig:hard}. Note that we use both the buggy and correct programs to infer the specifications. For each program, denoted as $prog$, \tool{} records executions across both sets of failing test cases $\mathbb{F}$ and set of related passing test cases $\mathbb{P}$. Typically, passing test cases $\mathbb{P}$ reflect the correct specification of a program while failing test cases $\mathbb{F}$ reflect the error specification of the program. Thus, \tool{} maintains the execution traces of the two test sets $\mathbb{F}$ and $\mathbb{P}$ separately to construct specifications for correct and error specification later on. 

To construct these specifications, \tool{} leverages \textsc{Daikon}~\cite{ernst2007daikon} to infer likely invariants based on the execution traces. \textsc{Daikon} first captures runtime values of variables at specific points in a program, such as the points at which a method is entered or exited, and then it uses a set of templates that satisfy the runtime values to infer likely invariants, which are properties that hold over all of the executions. We refer to the invariants inferred from the passing and failing test cases of a program $prog$ as $\mathcal{I}^{F}_{prog}$ and $\mathcal{I}^{P}_{prog}$, respectively. 

We use $\mathbb{B}$ and $\mathbb{G}$ to respectively denote the original buggy and correct (ground truth) versions of a program. To approximate the correct specification of $\mathbb{G}$, \tool{} infers invariants from the passing test cases on $\mathbb{B}$ and $\mathbb{G}$, denoted as $\mathcal{I}^{P}_{\mathbb{B}}$ and $\mathcal{I}^{P}_{\mathbb{G}}$ respectively. The correct specification $\mathcal{C}$ of $\mathbb{G}$ is then constructed by intersecting the two sets of invariants $\mathcal{I}^{P}_{\mathbb{B}}$ and $\mathcal{I}^{P}_{\mathbb{G}}$ and taking the resulting invariants as an approximation for the correct specification of $\mathbb{G}$. To approximate the error specification $\mathcal{E}$ in $\mathbb{B}$, \tool{} first infers invariants from the failing test cases on both $\mathbb{B}$ and $\mathbb{G}$, denoted as $\mathcal{I}^{F}_{\mathbb{B}}$ and $\mathcal{I}^{F}_{\mathbb{G}}$ respectively. The error specification $\mathcal{E}$ is then constructed by taking the invariants that are in $\mathcal{I}^{F}_{\mathbb{B}}$ but are not in $\mathcal{I}^{F}_{\mathbb{G}}$. In summary, the correct specification $\mathcal{C}$ represents the expected specification in $\mathbb{B}$ and $\mathbb{G}$, while the error specification $\mathcal{E}$ represents the behavioral difference of $\mathbb{B}$ compared to $\mathbb{G}$. 

\tool{} uses the constructed specifications $\mathcal{C}$ and $\mathcal{E}$ as inputs to its patch classifier, which we describe in Section~\ref{sec:classfication}, to identify overfitting patches. Note that the \tool{} classifier considers invariants inferred from all methods executed by a given test suite, rather than only invariants inferred from buggy methods (i.e., methods modified by human developers in the correct program) as in prior works~\cite{wang2020automated, yang2020exploring}. We discuss the effectiveness of the classifier using these two granularities in detail in Section~\ref{sec:eval}.

\subsubsection{Patch Classifier} \label{sec:classfication}
The patch classifier takes as input an APR-generated patch, the constructed specifications including correct specification $\mathcal{C}$ and error specification $\mathcal{E}$ to determine whether the patch is overfitting. 

\begin{algorithm}
	\KwIn{\begin{itemize}
	        \item $\mathcal{I}^{P}_{\mathbb{P}}$: invariant inferred from passing tests on $\mathbb{P}$
	        \item $\mathcal{I}^{F}_{\mathbb{P}}$: invariant inferred from failing tests on $\mathbb{P}$
			\item $\mathcal{I}^{P}_{\mathbb{B}}$: invariant inferred from passing tests on $\mathbb{B}$
			\item $\mathcal{I}^{F}_{\mathbb{B}}$: invariant inferred from failing tests on $\mathbb{B}$
			\item $\mathcal{I}^{P}_{\mathbb{G}}$: invariant inferred from passing tests on $\mathbb{G}$
			\item $\mathcal{I}^{F}_{\mathbb{G}}$: invariant inferred from failing tests on $\mathbb{G}$
		\end{itemize}
	}
	\KwOut{True: $\mathbb{P}$ is overfitting, False: Otherwise}
	\BlankLine
	$\mathcal{C} \gets \mathcal{I}^{P}_{\mathbb{G}} \cap \mathcal{I}^{P}_{\mathbb{B}}$ \Comment{Correct specification} \\
	\BlankLine
	$\mathcal{E} \gets \mathcal{I}^{F}_{\mathbb{B}} \backslash \mathcal{I}^{F}_{\mathbb{G}}$ \Comment{Error  specification} \\
	\BlankLine
	\ForEach{$inv$ in $\mathcal{C}$}
	{
		\If{$inv \notin \mathcal{I}^{P}_{\mathbb{P}}$}
		{
			\Return True \\ 
		}
		
	}
	
	\ForEach{$inv$ in $\mathcal{E}$}
	{
		\If{$inv \in \mathcal{I}^{F}_{\mathbb{P}}$}
		{
			\Return True \\ 
		}
		
	}
	\Return False \\
	\caption{Invariant-based Specification Inference. $\mathbb{B}$ is the original buggy program, $\mathbb{P}$ is an APR-patched program, and $\mathbb{G}$ is the correct/ground truth program by developers}
	\label{alg:main}
\end{algorithm}

Our approach to identifying patch correctness is based on two key observations:

\begin{observation} A patch should be considered overfitting if it violates any of the correct specifications described in $\mathcal{C}$.
\end{observation}
\begin{observation}
A patch should be considered overfitting if it maintains any of the error specifications described in $\mathcal{E}$.
\end{observation}

\par The above observations can be translated into the two following rules that allow \tool{} to determine whether an APR-patched program is overfitting. Consider $\mathbb{B}$ to be a buggy program and $\mathbb{G}$ to be the human-written correct version of the program. Let $\mathbb{P}$ be an APR-patched program to be assessed for overfitting, and $\mathcal{I}_{\mathbb{P}}$ be the set of invariants inferred from $\mathbb{P}$. A patch is considered overfitting if it satisfies either of the following conditions:
\begin{itemize}
    \item \textbf{Overfitting-1}: The patch violates the specifications representing correct specification $\mathcal{C}$ for $\mathbb{B}$ and $\mathbb{G}$. More formally, $\exists inv \in \mathcal{C}: inv \notin \mathcal{I}_{\mathbb{P}}$
    \item \textbf{Overfitting-2}: The patch maintains any error behaviors described in $\mathcal{E}$ for $\mathbb{B}$. More formally, $\exists inv \in \mathcal{E}: inv \in \mathcal{I}_{\mathbb{P}}$
\end{itemize}

In the \textit{Overfitting-1} rule, we consider any APR-patched program $\mathbb{P}$ to violate the correct specification if the set of invariants inferred from $\mathbb{P}$, denoted as $\mathcal{I}_{M}$, excludes any invariants that are in the correct specifications $\mathcal{C}$. This helps guard against cases where the patch deletes some functionalities of the original program and thus excludes the specifications corresponding to the functionalities. In the \textit{Overfitting-2} rule, any patch that still maintains an invariant representing error specification in the original buggy program $\mathbb{B}$ is considered overfitting. 

Note that, \tool{} needs to compare an invariant to another to determine whether a patch falls into either of the overfitting rules we defined above. \tool{} achieves this by comparing invariants syntactically and semantically. If two invariants are not syntactically the same, \tool{} leverages an SMT solver, i.e., \textsc{Z3}~\cite{de2008z3}, to determine if they are semantically equivalent. Generally, two logical formulae $\mathcal{A}$ and $\mathcal{B}$ are equivalent if $(\mathcal{A} \Rightarrow \mathcal{B}) \wedge (\mathcal{B} \Rightarrow \mathcal{A})$. For example, $a >= b$ and $b <= a$ are syntactically different but are determined to be semantically equivalent by \textsc{Z3}; \textsc{Z3} determines that the formulae $(a >= b \Rightarrow b <= a) \wedge (b <= a \Rightarrow a >= b)$ are satisfiable, and hence $a >= b$ is equivalent to $b <= a$.
\subsubsection{Optimization via test selection}
\par Our observation is that not all test cases are relevant to the bug at hand. Therefore, before running \textsc{Daikon}, \tool{} performs \textit{test selection}, as described in Algorithm~\ref{alg:testselect} to select a subset of passing test cases that are related to the bug to collect execution traces. This way, the reduced test suite helps \tool{} optimize for running time without compromising its accuracy.

\begin{algorithm}
	\KwIn{\begin{itemize}
			\item $\mathbb{P}$: program
			\item $\mathbb{P}$: set of modified methods
			\item $\mathbb{T}$: set of test cases
		\end{itemize}
	}
	\KwOut{Related tests}
	\BlankLine
	$\mathbb{R} \gets \emptyset$ \Comment{Set of related tests} \\
	\ForEach{$test$ in $\mathbb{T}$}{
	    $t \gets Coverage \text{(} \mathbb{P}, test \text{)} $ \Comment{Test coverage}
	    \If{\textit{t cover at least one method} $\in \mathbb{P}$  }{
	            $\mathbb{R} \gets \mathbb{R} \cup \{ test \}$
	    }

	}
	\Return $\mathbb{R}$
	\caption{Test selection}
	\label{alg:testselect}
\end{algorithm}

\subsection{Syntactic-based Patch classifier} \label{sec:softclassifier}

\begin{figure}[t]
    \centering
    \includegraphics[width=0.7\columnwidth]{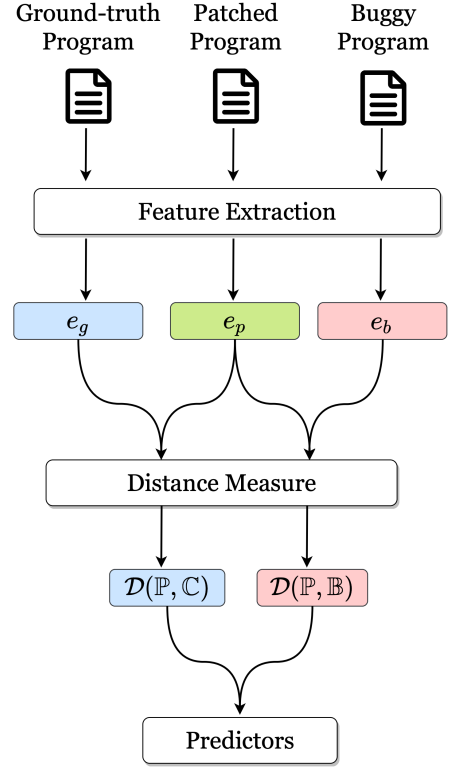}
    \caption{Model architecture of the syntactic classifier. $e_b$, $e_p$, and $e_c$ are representations of the buggy program, patched program, and ground truth program, respectively. $\mathcal{D}(\mathbb{P}, \mathbb{B})$, $\mathcal{D}(\mathbb{P}, \mathcal{G})$ are distances from patched program to buggy program and correct program}
    \label{fig:knowledge}
\end{figure}

In case \tool{} fails to reason about patch correctness via invariant-based specification inference (described in Section~\ref{sec:hardclassifier}), \tool{} resorts to estimating the probability that an APR-patched program is overfitting by measuring the syntactic proximity of the patch to the buggy and ground truth programs. Given an APR-patched program $\mathbb{P}$, \tool{} first measures the syntactic differences between $\mathbb{P}$ and the buggy program $\mathbb{B}$, denoted as $\mathcal{D}(\mathbb{P}, \mathbb{B})$, and between $\mathbb{P}$ and its ground truth program $\mathbb{G}$, denoted as $\mathcal{D}(\mathbb{P}, \mathbb{G})$. \tool{} employs a pre-trained language model, i.e., \textsc{CodeBERT}~\cite{feng2020codebert}, to extract syntactic features of these programs and then uses comparison functions~\cite{wang2017compare} as distance measures to identify syntactic differences between them. Finally, \tool{} uses a machine learning model to predict patch correctness. Figure~\ref{fig:knowledge} illustrates the classification pipeline of our syntactic-based classifier. Below, we explain each component of the pipeline in detail.

\subsubsection{Feature Extraction} 
The feature extraction layers aim to extract embedding vectors (a.k.a. features) that represent the syntactic information of buggy, patched, and correct programs. To achieve this, we utilize \textsc{CodeBERT}~\cite{feng2020codebert}, a powerful pre-trained model for general-purpose representations of source code that has demonstrated its effectiveness on various software engineering tasks~\cite{feng2020codebert, zhou2021assessing, nguyen2022vulcurator, le2022autopruner}. \textsc{CodeBERT} takes a code fragment as input and uses a tokenizer (i.e., the Roberta tokenizer) to tokenize the code into a sequence of tokens. It then passes the sequence through a pre-trained multi-layer bidirectional Transformer~\cite{vaswani2017attention} to obtain a corresponding numerical vector. Specifically, given a code fragment, \tool{} employs \textsc{CodeBERT} to represent the code fragment as the vector defined as follows:

\begin{equation}
e_{code}=\left\langle v_{1}, v_{2}, \ldots, v_{k}\right\rangle
\end{equation}
where $k = 768$ is the embedding dimension of \textsc{CodeBERT}. For convenience, we denote $e_b$, $e_p$, and $e_c$ as representations of the buggy program, patched program, and correct program, respectively.

\subsubsection{Distance Measure} 
The goal of the distance measure layers is to build the vectors that capture the syntactic differences between the APR-patched program, buggy program, and correct program. Inspired by prior works~\cite{tian2020evaluating, hoang2020cc2vec}, we leverage comparison functions~\cite{wang2017compare} to represent various types of syntactic differences. The distance measure layers take as input the embedding vectors of the buggy program, patched program, and correct program (denoted by $e_b$, $e_p$, and $e_c$, respectively) and output the vectors that represent the syntactic difference of the APR-patched program compared to the buggy program and correct program. These vectors are then concatenated to represent distance vectors, which have $2 \times k+2$ dimensions where $k$ is the dimension of code embeddings (i.e., 768). In this paper, we use four comparison functions, consisting of cosine similarity, Euclidean distance, element-wise subtraction, and multiplication. We briefly explain these comparison functions below.

\vspace{2mm}
\noindent \textbf{Element-wise subtraction.} We perform element-wise subtraction for the embedding vectors of the APR-patched program and the buggy program and the correct program as follows:
$$
e^{\mathrm{sub}}_{1} =e_{p} - e_{b}
$$

$$
e^{\mathrm{sub}}_{2} =e_{p} - e_{c}
$$

\vspace{2mm}
\noindent \textbf{Element-wise multiplication.} We perform element-wise multiplication for the embedding vectors of the APR-patched program and the buggy program and the correct program as follows:
$$
e^{\mathrm{mul}}_{1} =e_{p} \odot e_{b}
$$

$$
e^{\mathrm{mul}}_{2} =e_{p} \odot e_{c}
$$
where $\odot$ is the element-wise multiplication operator.
\vspace{2mm}

\noindent \textbf{Euclidean Distance.} We capture the distance between the embedding vectors of the APR-patched program and the buggy program and the correct program based on Euclidean distance as follows:

$$
e^{\mathrm{euc}}_{1} = \left\|e_{p}-e_{b}\right\|
$$

$$
e^{\mathrm{euc}}_{2} = \left\|e_{p}-e_{c}\right\|
$$
\vspace{2mm}
where $\left\|\cdot\right\|$ is the Frobenius norm.

\noindent \textbf{Cosine Similarity.} We capture the similarity between the embedding vectors of the APR-patched program and the buggy program and the correct program based on Cosine similarity as follows:

$$
e^{\mathrm{sim}}_{1} = \frac{e_{p} e_{b}}{\left\|e_{p}\right\|\left\|e_{b}\right\|}
$$

$$
e^{\mathrm{sim}}_{2} = \frac{e_{p} e_{c}}{\left\|e_{p}\right\|\left\|e_{c}\right\|}
$$
where $\left\|\cdot\right\|$ is the Frobenius norm.

\vspace{2mm}
\noindent \textbf{Distance vector.} Finally, we concatenated the vectors resulting from applying these three different comparison functions to represent the syntactic distances from the patched program to the buggy program and correct program as follows:
$$
\mathcal{D}(\mathbb{P}, \mathbb{B}) = \mathbf{e}^{\operatorname{sub}}_{1} \oplus \mathbf{e}^{\text {mul}}_{1} \oplus \mathbf{e}^{\mathrm{euc}}_{1} \oplus \mathbf{e}^{\mathrm{sim}}_{1}
$$

$$
\mathcal{D}(\mathbb{P}, \mathbb{C}) = \mathbf{e}^{\operatorname{sub}}_{2} \oplus \mathbf{e}^{\text {mul}}_{2} \oplus \mathbf{e}^{\mathrm{euc}}_{2} \oplus \mathbf{e}^{\mathrm{sim}}_{2}
$$
where $\oplus$ is the concatenation operation, $\mathcal{D}(\mathbb{P}, \mathbb{B})$ and $\mathcal{D}(\mathbb{P}, \mathbb{G})$ are distances from patched program to buggy program and correct program.
\subsubsection{Predictor} 
Given the above distance vectors, we leverage a machine-learning model to predict patch correctness from labeled data. Following the finding of Tian et al.~\cite{tian2020evaluating} that Logistic Regression applied to BERT embeddings yields the best performance in predicting patch correctness, we consider the Logistic Regression algorithm as our predictor. Logistic regression is a well-known machine learning (ML) algorithm that predicts patch correctness based on a linear transform and logistic loss function. 

\subsubsection{Correctness Prediction} \label{sec:finalscore}
\tool{} classifies a patch as correct or overfitting based on prediction score, i.e., the probability that a given patch is overfitting, produced by an ML-based predictor. Let 
$\mathcal{P}(m)$ denotes the prediction score of an APR-generated patch $m$. We determine the correctness of a given patch $m$ using the following formula:
$$
\text { correctness }(m)= \begin{cases}\text { correct } & \mathcal{P}(m) \leq \mathcal{T} \\ \text { overfitting } & \mathcal{P}(m) > \mathcal{T} \end{cases}
$$
where $\mathcal{T}$ is our classification threshold.

%% file: contents/evaluations.tex
\begin{table}[h]
    \centering
    \caption{Dataset for evaluating automated patch correctness assessment techniques}
    \resizebox{.5\textwidth}{!}{
    \begin{tabular}{l|c|c|c}
         \hline
         \textbf{Dataset} & \textbf{Correct patches} & \textbf{Overfitting patches} & \textbf{Total} \\
         \hline
         Xiong et al.~\cite{xiong2018identifying} & 30 & 109 & 139 \\  
         Wang et al.~\cite{wang2020automated} & 216 & 450 & 666 \\  
         \textsc{Defects4J}~\cite{just2014defects4j} & 223 & 0 & 223  \\ \hline
         \textbf{Final dataset} & \textbf{377} & \textbf{508} & \textbf{885} \\
         \hline
    \end{tabular}
    }
    \label{tab:data_stats1}
\end{table}


\begin{table}[h]
    \centering
    \caption{The statistics of evaluation and training dataset}
    \resizebox{0.5\textwidth}{!}{
    \begin{tabular}{l|c|c|c}
         \hline
         \textbf{Dataset} & \textbf{Correct patches} & \textbf{Overfitting patches} & \textbf{Total} \\
         \hline
         Training & 331 & 340 & 671  \\ 
         Validation & 16 & 59 & 75 \\ 
         Evaluation & 30 & 109 & 139 \\
         \hline
    \end{tabular}
    }
    \label{tab:data_stats2}
\end{table}

 In this section, we empirically evaluate \tool{} on a dataset of patches generated by well-known automated program repair techniques for bugs in large real-world Java programs. We discuss the dataset, experimental settings, and metrics in Section~\ref{sec:eval_settings}. Section~\ref{sec:eval_rq} lists our research questions, followed by our findings in Section~\ref{sec:eval_finding}. 
 
 \subsection{Experimental Settings} \label{sec:eval_settings} \subsubsection{Dataset} \label{sec:dataset} 
 
 To evaluate the effectiveness of automated patch correctness assessment (APCA) techniques, we have collected a dataset of APR-generated patches whose correctness labels were manually identified by independent developers and researchers. We used 220 patches released by Xiong et al.~\cite{xiong2018identifying} and 902 patches released by Wang et al.~\cite{wang2020automated}. Following previous works~\cite{xin2017identifying, tian2020evaluating}, we only consider patches from four widely-used projects in \textsc{Defects4J}: Chart, Time, Lang, and Math. This resulted in a dataset of 139 patches from Xiong et al.'s dataset and 666 patches from Wang et al.'s dataset. Additionally, to address data imbalance issues where very few APR-generated patches are labeled as correct, we supplemented the dataset with developer-written patches from the \textsc{Defects4J} dataset following~\cite{tian2020evaluating}. This resulted in a dataset of 1028 patches, including 469 correct patches and 559 overfitting patches. 
 
We consider 666 patches from Wang et al.'s dataset and 223 developer's patches from \textsc{Defects4J}~\cite{just2014defects4j} as the training and validation set and 139 patches from Xiong et al.~\cite{xiong2018identifying} as evaluation set following previous work~\cite{tian2020evaluating, xiong2018identifying, tian2020evaluating, ye2019automated}. Note that, there may be duplication between Wang et al. 's dataset, Defects4J's patches, and Xiong et al.'s dataset. To avoid data leakage, we removed the duplicated patches from the training and validation set. Particularly, we removed a patch if it is syntactically equivalent to a patch in the evaluation set. As a result, we obtained 746 (out of 889) patches for the training and validation phase. This included 347 correct patches and 399 overfitting patches. We use 90\% of these patches (90\% $\times$ 746 $=$ 671 patches) for training our learning model and the remaining 75 patches for validation.

Table~\ref{tab:data_stats1} shows the details of the patches considered in our experiments, and Table~\ref{tab:data_stats2} provides information on the characteristics of our training, validation, and evaluation datasets.

\subsubsection{Evaluation Metrics} \label{sec:metric}
By using the dataset described in Section~\ref{sec:dataset}, we assess the effectiveness of automated patch correctness assessment (APCA) techniques by comparing the labels produced by APCA versus the ground truth labels. Furthermore, we aim to assess how many patches an APCA technique produces that match that of the ground truth labels. Specifically, we use standard metrics of classification problems~\cite{tian2012information, tian2012identifying}, \textit{Recall} (Equation~\ref{equ:recall}), \textit{Precision} (Equation~\ref{equ:precision}), \textit{Accuracy} (Equation~\ref{equ:acc}), and  \textit{F1-score} (Equation~\ref{equ:F1-score}); they are defined by the following metrics:
\begin{itemize}
    \item \textbf{True Positive (TP)}: a generated patch is labeled as ``overfitting'' by both an APCA technique and the ground truth.
    \item \textbf{False Positive (FP)}: a generated patch is labeled as ``overfitting'' by an APCA technique but is labeled as ``correct'' by the ground truth.
    \item \textbf{True Negative (TN)}: a generated patch is labeled as ``correct'' by both an APCA technique and the ground truth.
    \item \textbf{False Negative (FN)}:  a generated patch is labelled as ``correct'' by an APCA technique, but is labelled as ``overfitting'' by the ground truth.
\end{itemize}
\begin{equation}
    \mathbf{Recall} = \dfrac{TP}{TP \text{ + } FN}
    \label{equ:recall}
\end{equation}
\begin{equation}
    \mathbf{Precision} = \dfrac{TP}{TP \text{ + } FP}
    \label{equ:precision}
\end{equation}

\begin{equation}
    \mathbf{Accuracy} = \dfrac{TP + TN}{TP + FP + TN + FN}
    \label{equ:acc}
\end{equation}

\begin{equation}
    \mathbf{F1} = \dfrac{2 \text{ x } Recall \text{ x } Precision}{(Precision \text{ + } Recall)}
    \label{equ:F1-score}
\end{equation}

Among these evaluation measures, \textit{Recall} verifies whether an approach can successfully classify overfitting patches. A higher \textit{Recall} is demanded by developers as we do not want to waste their efforts on analyzing a substantial number of overfitting patches~\cite{tao2014automatically}. Meanwhile, \textit{Precision} measures the proportion of discarded patches by an approach that is genuinely overfitting.  A higher \textit{Precision}  is desired by program repair research as we do not want to discard correct patches~\cite{ye2019automated}.

However, the comparison of APAC techniques that relies only on \textit{Recall} or \textit{Precision} may be incomplete. For example, 
an APAC technique can only consider patches as overfitting if it violates strict conditions (e.g., a high confidence value) to achieve a higher \textit{Precision}, which could result in a low \textit{Recall}. On the contrary, an approach can classify all patches as overfitting to achieve perfect \textit{Recall}, which results in low Precision. To address these issues, we consider \textit{F1-score} and \textit{Accuracy} as additional evaluation metrics to measure the performance of APAC techniques following previous studies~\cite{ye2021automated, lin2022context, zhou2023patchzero}. \textit{F1-score} seeks a balance between \textit{Recall} and \textit{Precision} while \textit{Accuracy} is the comprehensive evaluation of all TP, FP, TN, and FN.

Besides, we also consider Area Under the Curve ( denoted as \textbf{AUC}), which is defined as follows. 
\begin{equation}
    \mathbf{AUC} = \dfrac{S_{0} - n_{0}(n_{0} + 1)/2}{n_{0}n_{1}}
    \label{equ:auc}
\end{equation}
where $n_{0}$ and $n_{1}$ are the numbers of overfitting and correct patches, respectively, and $S_{0} = \Sigma{r_{i}}$, where $r_{i}$ is the rank of the $i^{th}$ overfitting patch in the descending list of output probability produced by each model. 

\textit{AUC} is a widely-used metric to evaluate the effectiveness of threshold-dependent classifiers~\cite{zhou2021finding, lo2009classification}. In our paper, \textit{AUC} is essential to compare the performance of syntactic-based classifiers. 

\subsubsection{Implementation Details} \label{sec:impl_details}
For \tool{}, we implement the proposed approach using Python programming language. For the \textsc{CodeBERT} model, we use HuggingFace's Transformers framework ~\footnote{https://huggingface.co/docs/transformers/index} as recommended by their authors in \textsc{CodeBERT}'s GitHub repository~\footnote{https://github.com/microsoft/CodeBERT}. With respect to the threshold $\mathcal{T}$ of the syntactic-based classifier, we set the default threshold at 0.975. To choose a classification threshold, we constraint the threshold to avoid filtering out any correct patches as following prior works~\cite{xiong2018identifying, tian2020evaluating}. Note that, we tune our classification threshold on an independent validation set (see details in Section~\ref{sec:dataset}) instead of the evaluation set as prior works~\cite{xiong2018identifying, tian2020evaluating} to avoid overfitting. We also investigate the impact of the threshold on the performance of \tool{} in Section~\ref{sec:eval}.

With respect to baseline techniques, we collect results of \ods{}, \textsc{PatchSim}, \textsc{Anti-patterns}, and \textsc{Bert + Lr} from prior works~\cite{ye2019automated, xiong2018identifying, tian2020evaluating}. For \textsc{DiffTGen} and \textsc{GT-Invariant}, we run their implementation to obtain their prediction for Xiong et al. dataset due to the lack of the result in the literature.

\subsection{Research Questions} \label{sec:eval_rq}

Our evaluation aims to answer these research questions:
\vspace{0.2cm}

\noindent \textbf{RQ1:} \textit{How effective is our approach to validate patches generated by automatic repair tools?} 

\noindent The research question concerns the ability of \tool{} for identifying overfitting patches generated by automated program repair techniques.
To demonstrate the value of our approach for automated patch correctness assessment tasks, we conduct an experiment in a dataset of 885 APR-generated patches (as described in Section~\ref{sec:dataset}) in terms of \textit{Precision}, \textit{Recall}, \textit{Accuracy}, and \textit{F1-score}. Then, we compare our approach to state-of-the-art baseline techniques, namely: 
\begin{itemize}
     \item  \rgt{}: Ye et al.~\cite{ye2021automated} proposed to use a testing procedure, named \underline{R}andom \underline{T}esting with \underline{G}round truth (\rgt{})~\cite{shamshiri2015automatically}, for APAC. Particularly, \rgt{} automatically generates tests based on developer-patched programs, which encodes the correct program behavior. And then if any automatically generated test fails on an APR-patched program, \rgt{} considers the program as \textit{overfitting}.  
    \item \ods{}: Ye et al.~\cite{ye2019automated} proposed \ods{}, an overfitting detection system. \ods{} builds machine learning classifiers based on 4,199 manually-crafted features for classifying overfitting patches;
    \item \textsc{Bert + Lr}: Tian et al.~\cite{tian2020evaluating} proposed a learning-based APAC technique that utilizes \textsc{Bert}~\cite{devlin2018BERT} and Logistic Regression to learn representations of code changes from historical data to predict the correctness of APR-generated patches. In this paper, we refer to their technique as \textsc{Bert + Lr};
    \item \textsc{PatchSim}: Xiong et al.~\cite{xiong2018identifying} proposed a dynamic APAC technique based on the similarity of execution trace similarity. In this paper, we refer to their technique as \textsc{PatchSim};
    \item \textsc{DiffTGen}: Xin et al.~\cite{xin2017identifying}
    proposed an APAC technique that identifies overfitting patches through test case generation. \textsc{DiffTGen} is the closest baseline related to our approach. Both \tool{} and \textsc{DiffTGen} assume the ground truth patches are available;
    \item \textsc{ \textsc{Anti-pattern}s}: In~\cite{tan2016anti}, the authors proposed seven generic categories of program transformation to detect overfitting patches. In this paper, we refer to their technique as \textsc{ \textsc{Anti-pattern}s};
    \item \textsc{GT-Invariant}: Recently, Yang and Yang~\cite{yang2020exploring} discovered that the majority of overfitting patches exhibit distinct runtime behaviors captured by the invariants generated by \textsc{GT-Invariant}~\cite{ernst2007daikon}. Building on this insight, Wang et al.~\cite{wang2020automated} propose a straightforward heuristic that considers an APR-generated patch as {\em overfitting} if any of its inferred invariants differ from those of the correct program. In this paper, we adopt their technique and refer to it as \textsc{GT-Invariant}.
\end{itemize}
\vspace{0mm}

\begin{table*}[t]
\centering
\caption{Comparison of the effectiveness of \tool{} with the state-of-the-art techniques. The bold numbers denotes the best result for Accuracy and F1-score.}\label{tab:comparison}
\resizebox{0.75\textwidth}{!}{
\begin{tabular}{l|cccc|cc|cc}
\hline
\textbf{Techniques} & {\textbf{TP}} & \textbf{FN}  & { \textbf{FP}} & \textbf{TN} & \textbf{Recall}                               & \textbf{Precision}                            & \textbf{Accuracy}                             & \textbf{F1-score}                                   \\ \hline
\textbf{\textsc{Anti-pattern}}                                            & { 27} & 82  & { 1}  & 29 & 0,25                                 & 0,96                                 & 0,40                                 & 0,39                                 \\
\textbf{\textsc{DiffTGen}}                                                & { 16} & 93  & { 0}  & 30 & 0,15                                 & 1,00 & 0,33                                 & 0,26                                 \\
\textbf{\textsc{PatchSim}}                                                & { 62} & 47  & { 0}  & 30 & 0,57                                 & 1,00                        & 0,66                                 & 0,73                                 \\
\textbf{GT-Invariant}                                            & { 59} & 50  & { 7}  & 23 & {0,54} & {0,89} & 0,59                                 & 0,67                                 \\
\textbf{\textsc{Bert + Lr}}                                         & { 43} & 66  & { 0}  & 30 & 0,39                                 & 1,00                        & 0,53                                 & 0,57                                 \\
\textbf{\ods{}}                                             & { 70} & 39  & { 5}  & 25 & 0,64                                 & 0,93                                 & 0,68                                 & 0,76                                 \\
\textbf{\rgt{}}                                             & { 70} & 39  & { 5}  & 25 & 0,64                                 & 0,93                                 & 0,68                                 & 0,76                                 \\\hline

\textbf{\tool{}}                                            & 86                        & 23  & 3                         & 27 & 0,79 & 0,97                                 & \textbf{0,81}&  \textbf{0,87}
\\ \hline
\end{tabular}
}
\end{table*}

\modify{
\noindent \textit{\textbf{RQ2:} How effective is our syntactic-based classifier?} 

This research question investigates the effectiveness of our syntactic-based classifier in assessing patch correctness. Toward this, we conduct experiments to answer two sub-questions:
\begin{itemize}
    \item  \textbf{RQ2.1:} \textit{How does our syntactic-based classifier compare to existing techniques?} In this research question, we compared our syntactic-based classifier to existing techniques, including \ods{} and \textsc{Bert+Lr} in terms of \textit{Precision}, \textit{Recall} \textit{Accuracy}, and \textit{F1-score} as RQ1. Besides, we also compare the performance of these techniques on \textit{\textit{AUC}}, a widely-used metric to evaluate the effectiveness of threshold-dependent classifiers. 
    \item  \textbf{RQ2.2:} \textit{How do our syntactic features compare to existing features?} In this research question, we investigate the effectiveness of our syntactic features extracted from \textsc{CodeBERT}, compared to syntactic features  extracted from existing methods, i.e., \ods{} and \textsc{BERT}.
\end{itemize}
}

\vspace{0mm}

\noindent \textbf{RQ3:} \textit{ How does the classification threshold affect the overall performance?}

\tool{} employs a classification threshold to determine whether a patch is overfitting, based on the prediction score of machine learning-based predictors. For the first two research questions, we set the classification threshold at 0.975, which yielded the highest precision on the validation dataset for \tool{}. In this research question, we investigate the impact of threshold sensitivity on \tool{}'s performance. To this end, we conduct experiments to address two sub-questions:
\begin{itemize}
\item \textbf{RQ3.1: } \textit{How does the classification threshold affect the overall performance of \tool{}?} We systematically set different values for this threshold and investigate how it affects the results of \tool{}.
\item \textbf{RQ 3.2: } \textit{How does threshold sensitivity affect the performance of our approach compared to other threshold-dependent techniques such as \textsc{PatchSim} or \ods{}?} This research question aims to investigate the impact of threshold sensitivity on the performance of \tool{} compared to existing techniques.
\end{itemize}

\vspace{0mm}

\vspace{0mm}

\noindent \textbf{RQ4: } \textit{Which components of \tool{} contribute to its performance?}

This research question aims to analyze the contribution of different components of \tool{} to its overall performance. Firstly, we investigate the impact of semantic and syntactic classifiers on \tool{}'s performance. Next, we examine the impact of design choices for each component, including the granularity of invariants and overfitting rules, on the performance of \tool{}. Specifically, we address three sub-questions as follows:

\begin{itemize}
    \item \textbf{RQ4.1:} \textit{How do semantic and syntactic classifiers affect the performance of our approach?} \tool{} contains two main components: semantic and syntactic classifiers. In this research question, we perform an ablation study by dropping each classifier to evaluate the contribution of each classifier to \tool{}'s performance. 
    
    \item \textbf{RQ4.2:} \textit{Does using invariants inferred from executed methods improve the performance of \tool{} compared to using invariants inferred from buggy methods only?} By default, \tool{} considers invariants inferred from all methods executed by a given test suite, rather than only using invariants inferred from buggy methods as done by prior works~\cite{wang2020automated, yang2020exploring}. In this research question, we investigate the effectiveness of these two granularities.
    
    \item \textbf{RQ4.3:} \textit{How do overfitting rules affect the performance of our semantic classifier?} By default, our semantic classifier uses a combination of the \emph{Overfitting-1} and \emph{Overfitting-2} rules described in Section~\ref{sec:classfication} to identify overfitting patches. In this research question, we compare these two overfitting rules individually to evaluate their impact on \tool{}'s effectiveness.    
\end{itemize}

\subsection{Findings}\label{sec:eval_finding}
\subsubsection{RQ1: Effectiveness} \label{sec:effectiveness}
We report the comparison of our approach, \tool{}
against baseline techniques consisting of \rgt{}~\cite{ye2021automated}, \ods{}~\cite{ye2019automated}, \textsc{Bert}+LR~\cite{tian2020evaluating}, \textsc{PatchSim}~\cite{xiong2018identifying}, \textsc{DiffTGen}~\cite{xin2017identifying}, \textsc{ \textsc{Anti-pattern}s}~\cite{tan2016anti}, \textsc{GT-Invariant}~\cite{yang2020exploring} on 139 APR-generated patches collected by Xiong et al.~\cite{xiong2018identifying}. 
Table~\ref{tab:comparison} presents the detailed results with respect to evaluation metrics given in Section~\ref{sec:metric}, including \textit{Recall}, \textit{Precision}, \textit{Accuracy}, and \textit{F1-score}. We highlight the best result for each evaluation metric as bold numbers. 
The bold red number denotes the metrics for which the \tool{} shows the highest results among the techniques. 

Overall, \tool{} successfully identifies correctly 86 out of 109 overfitting patches and misclassified 3 out of 30 correct patches, equivalent to scores of 0.79, 0.97, 0.81, and 0.87 in terms of \textit{Recall}, \textit{Precision}, \textit{Accuracy}, and \textit{F1-score}, respectively. 
This implies that \tool{} outperforms all baselines in \textit{Recall}, \textit{Accuracy}, and \textit{F1-score} and obtains a good \textit{Precision} of 0.97. We present more details below.

\vspace{1mm}

\noindent \textbf{Accuracy.} Table~\ref{tab:comparison} shows that \tool{} correctly identifies 86 out of 109 overfitting patches and 27 out of 30 correct patches, resulting in an \textit{Accuracy} of 0.81. This indicates that \tool{} outperforms the best baselines (\ods{} and \rgt{}) by 19\% and shows improvements of 23\% to 146\% compared to the other baselines.

\vspace{1mm}

\noindent \textbf{F1-score}
\noindent \textbf{F1-score.} \tool{} outperforms the two best baselines (i.e., \ods{} and \rgt{}) by 14\%. Specifically, \tool{} outperforms \textsc{Bert}+LR, \textsc{PatchSim}, \textsc{DiffTGen}, \textsc{Anti-patterns}, and \textsc{GT-Invariant} by 54\%, 20\%, 239\%, 120\%, and 29\%, respectively. This is mainly because \tool{} successfully identifies 79\% of overfitting patches, whereas the best baselines filter out only 64\% of overfitting patches while still maintaining an acceptable precision of 0.97.

\vspace{1mm}

\noindent \textbf{Recall.} With respect to \textit{Recall}, \tool{} achieves improvements of 23\% (0.79 vs. 0.64) compared to the best baselines (i.e., \ods{} and \rgt{}). Specifically, \tool{} outperforms \rgt{}, \ods{}, \textsc{Bert}+LR, \textsc{PatchSim}, \textsc{DiffTGen}, \textsc{Anti-patterns}, and \textsc{GT-Invariant} by 23\%, 23\%, 100\%, 39\%, 438\%, 219\%, and 46\%, respectively. This is mainly because \tool{} leverages both syntactic and semantic reasoning while other techniques consider only syntax or semantics alone.

\vspace{1mm}

\noindent \textbf{Precision.} In terms of \textit{Precision}, \tool{} achieves a score of 0.97, outperforming \textsc{Anti-patterns}, \ods{}, \rgt{}, and \textsc{GT-Invariant}, which have \textit{Precision} scores of 0.96, 0.93, 0.93, and 0.89, respectively. However, \tool{} slightly underperforms \textsc{Bert}+LR and \textsc{PatchSim} in terms of \textit{Precision}. This may be because \textsc{Bert}+LR and \textsc{PatchSim} avoid filtering out correct patches by directly tuning the threshold of their classifier on the evaluation set, which could lead to overfitting on the set. In contrast, we tune our classification threshold on an independent validation set (as presented in Section~\ref{sec:impl_details}) to avoid overfitting, which leads to lower precision than \textsc{Bert}+LR and \textsc{PatchSim} on the evaluation set. Meanwhile, \textsc{DiffTGen} has a perfect \textit{Precision} (i.e., 1.0), but it is much less effective in filtering out overfitting patches, as reflected by its low \textit{Recall} of 0.25\vspace{1mm}

\begin{figure}[ht]
    \centering
    \includegraphics[width=0.7\columnwidth]{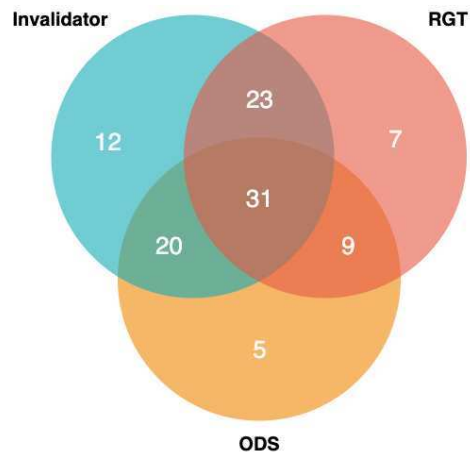}
    \caption{Intersection on the correctly classifier overfitting patches by \tool{}, \ods{} and \rgt{}}
    \label{fig:tp_venn}
\end{figure}

\noindent \textbf{Complementarity with \ods{} and \rgt{}.} We also perform a detailed analysis on the overfitting patches correctly classified by \tool{}, \ods{}, and \rgt{}. Figure 5 shows the intersection of their correctly classified overfitting patches. We can see that these techniques only detected 31/109 overfitting patches together, accounting for less than 40\% of the overfitting patches correctly classified by each technique. Meanwhile, \tool{}, \rgt{}, and \ods{} individually detect 10, 7, and 5 overfitting patches that are not detected by one another, respectively. More interestingly, the overfitting patches correctly classified by the three techniques cover most of the overfitting patches (107/109). These results suggest that the three techniques are complementary and can be used together to obtain a better patch correctness assessment.

\begin{figure}[t]
\centering
\begin{subfigure}[H]{0.4\textwidth}
\includegraphics[width=\textwidth]{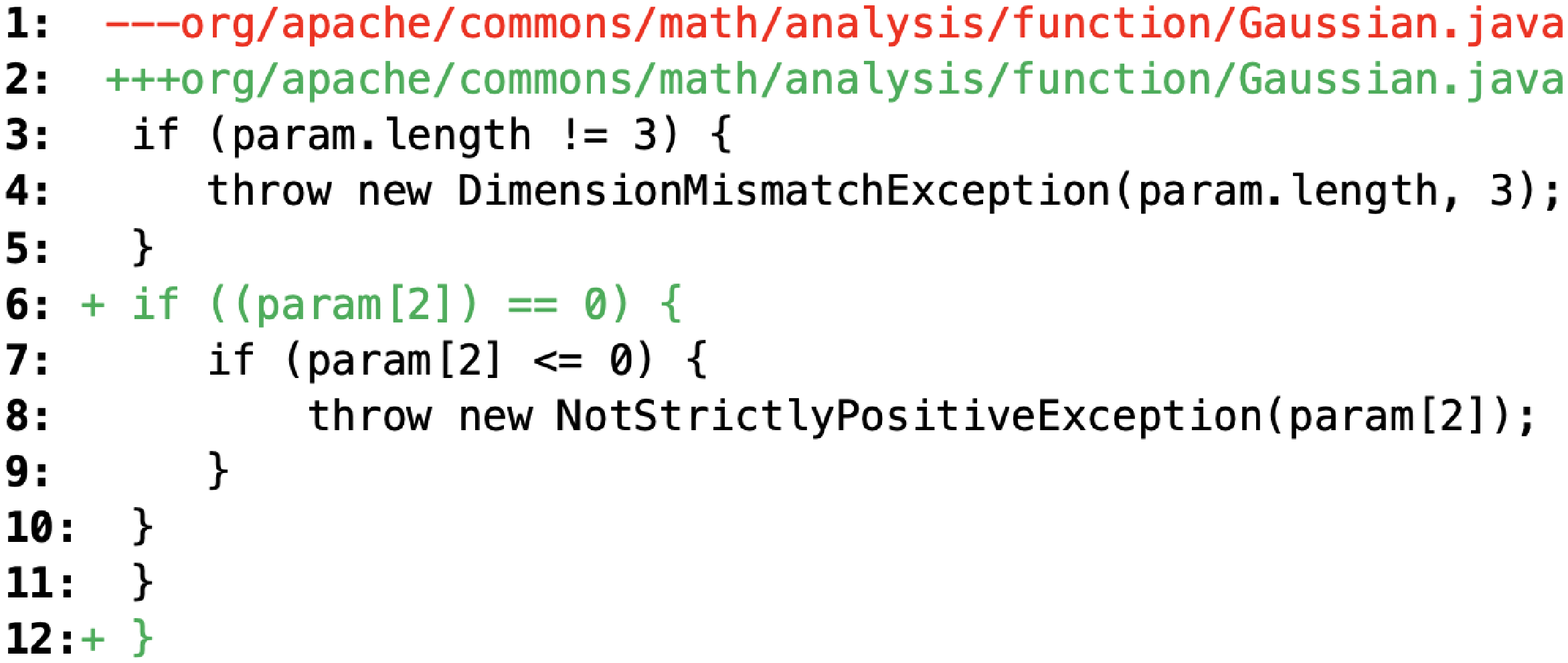}
\hfill
\caption{An overfitting patch generated by Nopol~\cite{xuan2016nopol} \label{fig:quali_patch}}
\end{subfigure}

\vspace{0mm}

\begin{subfigure}[H]{0.5\textwidth}
\includegraphics[width=\textwidth]{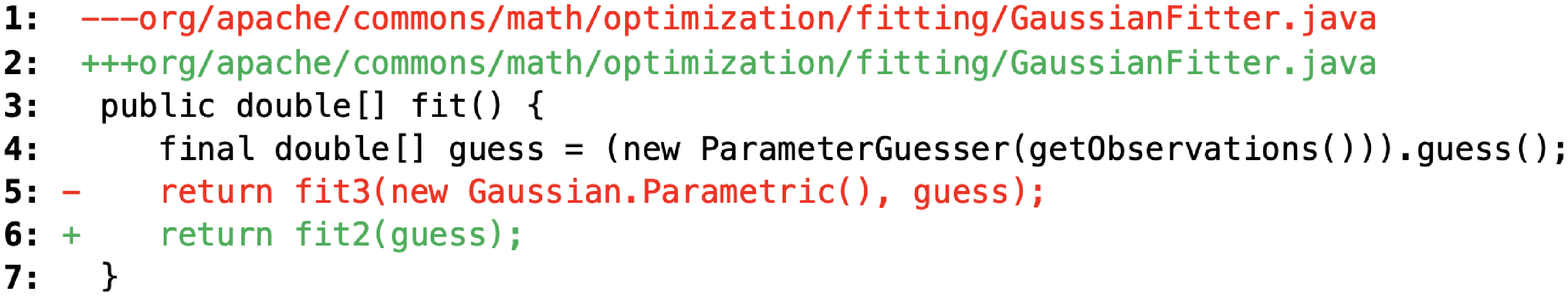}
\caption{The correct patch written by human developers \label{fig:quali_human}}
\end{subfigure}
\caption{An overfitting patch generated by Nopol and the human-written patch for Math-58}
\label{fig:quali}
\end{figure}

\vspace{0mm}

\noindent \textbf{Case study of unique overfitting patches.} To  provide further insights into our approach, we manually analyzed unique overfitting patches that can be detected with the help of the novel techniques in \tool{}.
In Figure~\ref{fig:quali}, we
present an example of an overfitting patch generated for the bug Math-58, which is detected as overfitting by \tool{} but not \rgt{} and \ods{}. 
In the bug, the \texttt{fit()} method (line 3 in Figure~\ref{fig:quali_human}) is utilized to fit a Gaussian function to the observed points. Ideally, the method should ideally catch the exceptions of observed points having a negative standard deviation and return NaN values. To achieve this, the method must call method \texttt{fit2()} to initialize a new Gaussian function and catch the exceptions before calling method \texttt{fit3()} to fit the Gaussian function. 
However, in the buggy version, \texttt{fit()} directly initializes a new Gaussian function and calls \texttt{fit3()} (line 5 in Figure~\ref{fig:quali_human}), which results in the buggy version missing the observed points having a negative standard deviation and throwing \texttt{NotStrictlyPositiveException}. As we can see in Figure~\ref{fig:quali_patch}, Nopol fixes the bug by adding the condition \texttt{param[2] == 0} (line 6), which ensures that the \texttt{NotStrictlyPositiveException} (line 8) is unreachable when the observed points have a negative standard deviation. This leads to the failing test case being plausibly passed, but the program is still incorrect. 
However, as it is no longer possible to trigger this error, \rgt{}, which relies on test case generation, fails to detect the different behaviors between the overfitting patch and the correct patch. 
In contrast, \tool{}, which relies on program invariants, still can correctly detect the overfitting patch. Indeed, in both buggy program and Nopol's patched program, \tool{} found the invariant \texttt{f.getClass() == Gaussian\$Parametric.class} at the entry point of method \texttt{fit3()}, indicating that the Gaussian function is directly initialized in method \texttt{fit()}. Meanwhile, the Gaussian function should be initialized in \texttt{fit2()} reflected by an invariant of the developer-patched program \texttt{f.getClass() == GaussianFitter\$1.class} at the entry point of method \texttt{fit3()}. We can see that Nopol's patch satisfies our Overfitting-2 rule, i.e., maintaining error behavior. Therefore, \tool{} can correctly classify the patch as overfitting. Another example can be seen in Section~\ref{sec:motivation}, in which an overfitting patch generated by Kali~\cite{qi2014strength} cannot be detected by \rgt{}, but can be detected by \tool{} as the patch violates our Overfitting-1 rule, i.e., it violates correct behavior.

\begin{tcolorbox}
\textbf{Answers to RQ1:} 
\tool{} yields very promising performance on assessing the correctness of APR-generated patches (\textit{Accuracy} at 0.81 and \textit{F1-score} at 0.87) and outperforms the best baseline by 19\% and 14\% in terms of \textit{Accuracy} and \textit{F1-score}, respectively. Besides, the complementary use of the three best-performing techniques can cover 107/109 overfitting patches. 
\end{tcolorbox}

\modify{
\subsubsection{RQ2: Effectiveness of syntactic-based classifier}
\begin{table*}[]
\centering
\caption{Comparison of the effectiveness of \tool{}'s syntactic-based classifier with the state-of-the-art techniques. The bold numbers denotes the best result for Accuracy, F1-score and AUC.} \label{tab:syntactic_model}
\resizebox{0.8\textwidth}{!}{
\begin{tabular}{|l|cccc|cc|cc|c|}
\hline
\textbf{Techniques} & {\textbf{TP}} & {\textbf{FN}} & {\textbf{FP}} & {\textbf{TN}} & {\textbf{Recall}} & {\textbf{Precision}} & {\textbf{Accuracy}} & {\textbf{F1-score}} & \textbf{AUC
} \\ \hline
\textbf{\textsc{Bert+Lr}} & 43 & 66 & 0 & 30 & 0.39 & 1.00 & 0.53 & 0.57 & 0.77    \\
\textbf{\ods{}} & 70 & 39 & 5 & 25 & 0.64 & 0.93 & 0.68 & 0.73 & 0.84    \\ \hline
\textbf{$\mathbf{\tool{}}_{Syn}$}  & 74 & 35 & 3 & 27 & 0.68 & 0.96 & \textbf{0.73} & \textbf{0.80} & \textbf{0.89} \\ \hline       
\end{tabular}
}
\end{table*}

\begin{table*}[]
\centering
\caption{Comparison of the effectiveness of \textsc{CodeBERT} features with \ods{} and \textsc{BERT} features. The bold numbers denotes the best result for Accuracy. F1-score and AUC.} \label{tab:syntactic_features}
\resizebox{0.8\textwidth}{!}{
\begin{tabular}{|l|l|cccc|cc|cc|c|}
\hline
\textbf{ground truth} & \textbf{Techniques} & {\textbf{TP}} & {\textbf{FN}} & {\textbf{FP}} & {\textbf{TN}} & {\textbf{Recall}} & {\textbf{Precision}} & {\textbf{Accuracy}} & {\textbf{F1-score}} & \textbf{AUC
} \\ \hline
&$\mathbf{BERT}_{wo-gt}$ & 33 & 76 & 2  & 28 & 0.30   & 0.94      & 0.44     & 0.46 & 0.71 \\ 
No &$\mathbf{ODS}_{wo-gt}$ & 25 & 84 & 0  & 30 & 0.23   & 1.00      & 0.40     & 0.37 & 0.77 \\ 
&$\mathbf{CodeBERT}_{wo-gt}$ & 36 & 73 & 2  & 28 & 0.33   & 0.95      & 0.46     & 0.49 & 0.83 \\ \hline

&$\mathbf{BERT}_{gt}$ & 68 & 41 & 5    & 25   & 0.66 & 0.92 & 0.69 & 0.77 & 0.83 \\ 
Yes &$\mathbf{ODS}_{gt}$ & 30 & 79 & 0 & 30   & 0.8 & 0.94 & 0.43 & 0.43 & 0.81 \\ 
&$\mathbf{CodeBERT}_{gt}$ & 74 & 35 & 3 & 27 & 0.68 & 0.96 & \textbf{0.73} & \textbf{0.77} & \textbf{0.89} \\ \hline
\end{tabular}
}
\end{table*}

\noindent \textbf{[RQ2.1: Our syntactic-based classifier vs. existing techniques]}

\noindent In this sub-question, we compare the performance of our syntactic-based classifier with two existing learning-based APAC techniques: \ods{} and \textsc{Bert}+LR. Table~\ref{tab:syntactic_model} presents the effectiveness of our approach and two baselines on six evaluation metrics including \textit{Accuracy}, \textit{F1-score}, and \textit{AUC}. The experimental results demonstrate that \tool{} significantly outperforms two baselines over six evaluation metrics. Particularly, \tool{} yields an \textit{Accuracy} of 0.73 and \textit{F1-score} of 0.80, outperforming the best baseline, i.e., \ods{}, by 6\% and 5\%, respectively. Note that \ods{} requires manual efforts to extract hand-crafted features while our patch classifier automatically extracts features based on labeled datasets. Compared to \textsc{Bert}+LR, which also uses automatically-extracted features, our syntactic classifier shows substantial improvement of 38\% and 41\% in terms of \textit{Accuracy} and \textit{F1-score}, respectively. Moreover, \tool{} also improves \ods{} and \textsc{Bert}+LR by 6\% and 16\% over AUC, indicating that our syntactic classifier has a better discriminative capability than existing techniques regardless of thresholds. 

\begin{tcolorbox}
\textbf{Answers to RQ2.1:} 
Our syntactic-based classifier significantly outperforms existing techniques over all evaluation metrics. Notably, our classifier also improves the best baseline by 6\% in terms of AUC, indicating it is more effective than existing techniques regardless of thresholds. 
\end{tcolorbox}

\noindent \textbf{[RQ2.2: Our syntactic features vs. existing syntactic features]}

\noindent In this sub-question, we compare the performance of our syntactic features extracted from CodeBERT with existing ones extracted from ODS and BERT regarding our syntactic-based classifier. To ease our presentation, we refer to the features  as CodeBERT’s, ODS’s, and BERT’s features, respectively. Table~\ref{tab:syntactic_features} presents the effectiveness of six variants of the syntactic-based classifier using three syntactic features: ODS’s, BERT’s, and CodeBERT’s features with and without ground truth knowledge.
The evaluation results showed that \textsc{CodeBERT}'s features significantly outperform \ods{}'s and \textsc{BERT}'s features. Particularly, with ground truth knowledge, \textsc{BERT}'s features show an improvement of 9\%, 8\%, and 7\% regarding \textit{Accuracy}, \textit{F1-score}, and \textit{AUC}, respectively. Meanwhile, the improvements without ground truth knowledge are 5\%, 7\%, and 17\%. Besides, we also can see that our classifier with ground truth knowledge improves the variants without the knowledge regardless of syntactic features over three metrics: \textit{Accuracy}, \textit{F1-score}, and \textit{AUC}. The improvement is especially substantial regarding threshold-dependent techniques, i.e., \textit{Accuracy} and \textit{F1-score}. These results indicate the advantage of adding ground truth knowledge for syntactic-based classifiers.

\begin{tcolorbox}
\textbf{Answers to RQ2.2:} 
CodeBERT's features are the most suitable features for our syntactic-based classifier. Besides, ground truth knowledge is helpful for syntactic-based classifiers.
\end{tcolorbox}
}

\subsubsection{Threshold Sensitivity}

\begin{figure}
     \begin{subfigure}[b]{0.5\textwidth}
         \centering
         \includegraphics[width=\textwidth]{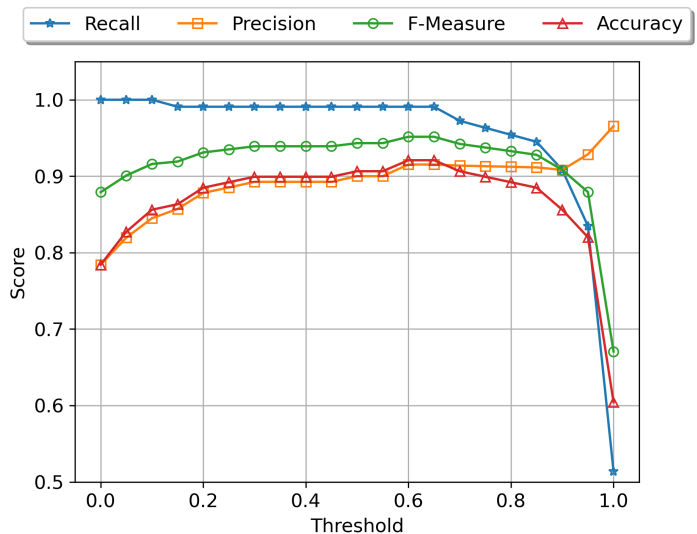}
     \end{subfigure}
        \caption{The performance of \tool{} with different classification thresholds on the evaluation set}
        \label{fig:threshold}
\end{figure}

\noindent \textbf{[RQ3.1: The impact of threshold sensitivity on the performance of \tool{}]} 

\noindent Recall that \tool{} uses a threshold, which ranges from 0 to 1, to classify whether a patch is overfitting based on a prediction score produced by Machine Learning predictors as defined in Section~\ref{sec:finalscore}. In this sub-question, we investigate the performance of \tool{} in terms of \textit{Recall}, \textit{Precision}, \textit{F1-score}, and \textit{Accuracy} with different classification thresholds in range (0, 1). The impact of the classification threshold on the performance of our approach is illustrated in Figure~\ref{fig:threshold}. 

We can see that the \textit{Recall} holds steady at around \textit{1.0} when the classification thresholds are in the range of (0.0, 0.65), then slightly decreases to about 0.94 (at the threshold of 0.85) before dropping to about \textit{0.53} at the maximum threshold of 1.0. On the contrary, as the classification threshold increases, the \textit{Precision} gradually increases from 0.79 to 0.97. \modify{Notably, \tool{}’s precision is always higher than 0.8 and around 0.9 at most of the thresholds. These results indicate that the assessment of \tool{} is reliable.}

With respect to \textit{Accuracy} and \textit{F1-score}, the performance of \tool{} shares a similar trend on these metrics according to the variation of the classification threshold. In detail, \textit{Accuracy} and \textit{F1-score} consistently increase from 0.79 and 0.89 to 0.92 and 0.95, respectively, when the threshold increases from 0.0 to about 0.6.  Then, these metrics slightly decrease to 0.84 of \textit{Accuracy} and 0.89 of \textit{F1-score} at the threshold of 0.9 before dropping to below 0.7 at the maximum threshold of 1.0. 

In summary, our results suggest that the classification threshold has a limited impact on the \textit{F1-score} and \textit{Accuracy}, despite its influence on \textit{Recall} and \textit{Precision}. Practitioners and researchers can therefore select a threshold that aligns with their needs, without compromising the discriminative ability of APAC techniques, as reflected by \textit{F1-score} and \textit{Accuracy}. 

\begin{tcolorbox}
\textbf{Answers to RQ3.1:} 
Despite the change of \textit{Precision} and \textit{Recall}, \tool{} still achieves promising overall performance, i.e., \textit{F1-score} and \textit{Accuracy} at above 0.8, over a large range of classification threshold, i.e., (0.1 - 0.9), on both validation and evaluation set. 
\end{tcolorbox}

\vspace{0mm}
\begin{figure}
     \centering
     \begin{subfigure}[b]{0.5\textwidth}
         \centering
         \includegraphics[width=\textwidth]{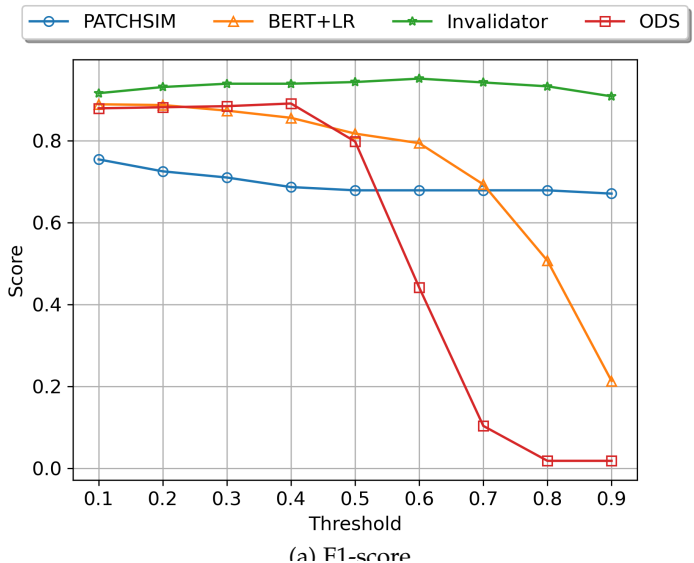}
         \caption{F1-score}
         \label{fig:threshold_F1-score}
     \end{subfigure}
     \hfill
     \begin{subfigure}[b]{0.5\textwidth}
         \centering
         \includegraphics[width=\textwidth]{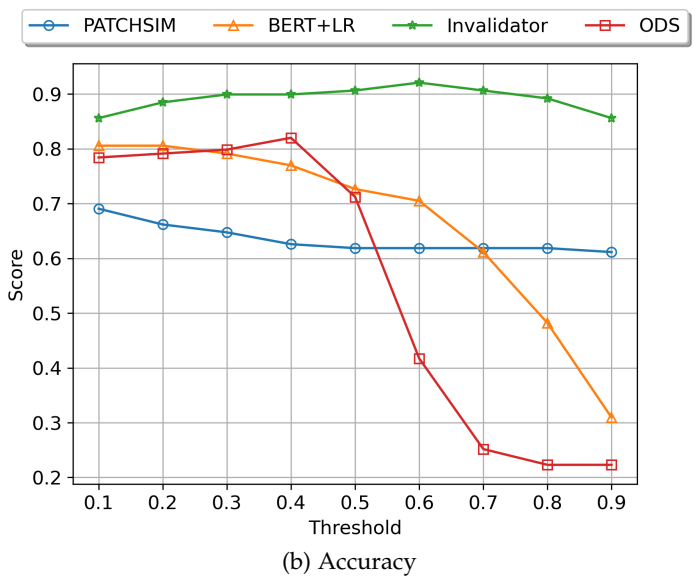}
         \caption{Accuracy}
         \label{fig:threshold_acc}
     \end{subfigure}
        \caption{The performance of \tool{}, \ods{}, \textsc{Bert}+LR and \textsc{PatchSim} with different classification thresholds}
        \label{fig:threshold_compare}
\end{figure}

\noindent \textbf{[RQ3.2: \tool{} vs. Existing threshold-dependence techniques]}

\noindent In this sub-question, we compare the performance, reflected by \textit{F1-score} and \textit{Accuracy)}, of four threshold-dependence techniques consisting of \tool{}, \ods{}~\cite{ye2019automated}, \textsc{Bert}+LR~\cite{tian2020evaluating} and \textsc{PatchSim}~\cite{xiong2018identifying} with nine different thresholds in the range of (0.1, 0.9). The impact of the classification threshold on the performance of threshold-dependence techniques is illustrated in Figure~\ref{fig:threshold_compare}. The results yield two main findings. First, the classification threshold has a limited impact on the performance of \tool{} and \textsc{PatchSim}. Meanwhile, \textsc{Bert}+LR and \ods{} only achieve good performance in the threshold range of (0.1, 0.4) before witnessing a significant decrease of both \textit{F1-score} and \textit{Accuracy} when the threshold increases from 0.4 to 0.9. The finding indicates that \tool{} and \textsc{PatchSim} are more stable than \textsc{Bert}+LR and \ods{} with respect to the variation of classification threshold. Second, \tool{}, with an arbitrary threshold, performs better than the best result of each baseline. The finding indicates that \tool{} is the most effective technique among threshold-dependence APAC approaches. 

\begin{tcolorbox}
\textbf{Answers to RQ3.2:} 
\tool{} is the most effective and stable technique among threshold-dependence APAC approaches. 
\end{tcolorbox}

\subsubsection{Ablation Study}

\begin{table*}[]
\centering
\caption{Ablation Study. The $\tool{}_{Syn}$ and $\tool{}_{Sem}$ denotes \tool{}'s syntactic and semantic classifiers, respectively. The bold numbers denotes the better result in each evaluation metric} \label{tab:ablation}
\resizebox{0.8\textwidth}{!}{
\begin{tabular}{|l|cccc|cccc|}
\hline
\textbf{Techniques} & {\textbf{TP}} & {\textbf{FN}} & {\textbf{FP}} & {\textbf{TN}} & {\textbf{Recall}} & {\textbf{Precision}} & {\textbf{Accuracy}} & {\textbf{F1-score}} \\ \hline
\textbf{\tool{}}                              & {86}          & {23}          & {3}           & {27}          & \textbf{0,79}            & \textbf{0,97}               & \textbf{0,81} & \textbf{0.87} \\ \hline       

-w/o $\mathbf{\tool{}}_{Syn}$                           & {56}          & {53}          & {2}           & {28}          & {0,51}            & \textbf{0,97}               & {0,60}   & 0.67    \\
-w/o $\tool{}_{Sem}$                           & {74}          & {35}          & {3}           & {27}          & {0,68}            & 0,96               & {0,73}   & 0.80    \\ \hline

\end{tabular}
}
\end{table*}

\modify{
\noindent \textbf{[RQ4.1: The impact of semantic-based and syntactic-based classifiers on the performance of \tool{}]} In this experiment, we evaluate the relative contribution of  \tool{}'s semantic versus structural classifier for patch correctness assessment.
Table~\ref{tab:ablation} shows the results of our experiments.
$\text{\tool{}}_{sem}$,  $\text{\tool{}}_{syn}$ refer to semantic and syntactic-based classifiers, respectively. In the ablation study, we can observe that \tool{} without these classifiers suffer from different degrees of performance loss. Specifically, removing $\text{\tool{}}_{syn}$ leads to a decrease of 26\% and 23\% in terms of Accuracy and F1-score; meanwhile without $\text{\tool{}}_{sem}$, \tool{}'s performance also drops by 11\% and 8\%, respectively. Also, we can see that our syntactic-based classifier shows a better performance than our semantic-based classifier. This is mainly because our semantic-based classifier can only detect 56 overfitting patches compared to 74 of our syntactic-based classifiers. One potential reason behind the phenomena is that our semantic-based classifier depends on our current test suite, which may be an incomplete and invariant generator, i.e., Daikon. Therefore, though our semantic-based classifier can reveal hidden behavior differences between the APR-patched and ground truth programs to detect overfitting patches, its effectiveness can still be bounded by the abovementioned factors.
However, the semantic-based classifier is still important for our approach to dealing with the threshold sensitivity of syntactic-based classifiers. Indeed, our semantic-based classifier is threshold-independent, allowing its performance to be considered a lower bound for the performance of \tool{}. Therefore, \tool{} still can work well with a strict classification threshold, making \tool{} become the most stable technique among threshold-dependence APAC approaches, as we can see in the RQ 3.2. These results suggest that both semantic and syntactic-based classifiers are essential for the performance of \tool{}.

\begin{tcolorbox}
\textbf{Answers to RQ4.1:} 
Our ablation study shows that both semantic and syntactic-based classifiers contribute to the effectiveness of \tool{}.
\end{tcolorbox}

}

\noindent \textbf{[RQ4.2: The impact of invariant granularity on the performance of \tool{}]} 

\begin{table}[h]
\caption{Overall performance of \textit{\tool{}'s semantic classifier} with different invariant granularity: buggy methods and executed methods. The bold numbers denotes the better result in each evaluation metric} \label{tab:invariant_granularity}
\resizebox{0.5\textwidth}{!}{
\begin{tabular}{
|l| 
c 
c| 
c 
c|}

\hline
\textbf{Granularity} & {\textbf{Recall}} & {\textbf{Precision}} & {\textbf{F1-score}}   & \textbf{Accuracy} \\
\hline
Buggy methods                                            & {0,35}            & {0,95}               & {0,51}          & 0,47                                                          \\ \hline
Executed methods                                          & {\textbf{0,51}}   & {\textbf{0,97}}      & {\textbf{0,67}} & \textbf{0,60}                                              
\\ \hline
\end{tabular}}
\end{table}

\noindent In this sub-question, we investigate the performance of our semantic classifier with invariant inferred from two different granularities: buggy methods and executed methods, i.e., methods executed by test cases. As shown in Table~\ref{tab:invariant_granularity}, the invariants inferred from executed methods can boost the performance of our semantic classifier in APAC by 28\% at \textit{Accuracy} and 31\% at \textit{F1-score}. The key reason for the improvement is that behavioral differences between APR-generated patches and correct patches exist in methods called by a statement of buggy methods. Hence, the supplement of invariant inferred from all executed methods helps our semantic classifier to detect more overfitting patches.

\begin{tcolorbox}
\textbf{Answers to RQ4.2:} 
The supplement of invariant inferred from all executed methods helps our semantic classifier boost the performance by 31\% at \textit{Accuracy} and 35\% at \textit{F1-score}
\end{tcolorbox}

\vspace{0mm}

\noindent \textbf{[RQ4.3: The impact of different overfitting rules on the performance of \tool{}]}

\noindent In this sub-question, we investigate the impact of each overfitting rule on the performance of our semantic classifier. 

\begin{figure}[h]
    \centering
    \includegraphics[width=0.65\columnwidth]{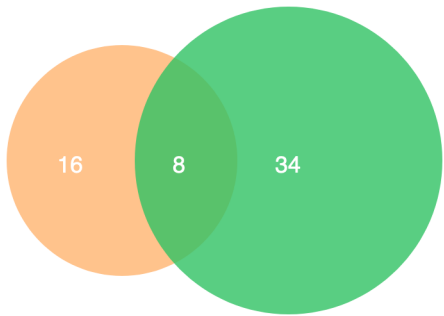}
    \caption{The impact of overfitting rules on the performance of \textit{\tool{}'s semantic classifier}}
    \label{fig:overfitting_rules}
\end{figure}

\noindent As shown in Figure~\ref{fig:overfitting_rules} the \textit{Overfitting-1} and \textit{Overfitting-2} contributes 24 and 42 overfitting patches, respectively, among 56 patches detected by our semantic classifier. Moreover, there are 8 overfitting patches violating both overfitting rules. The results indicate that \textit{Overfitting-2} rule contributes to our semantic classifier much more than \textit{Overfitting-1}.

\begin{tcolorbox}
\textbf{Answers to RQ4.3:} 
\textit{Overfitting-2} rule contributes to \tool{} much more than \textit{Overfitting-1} (42 vs. 24 overfitting patches).
\end{tcolorbox}

\vspace{0mm}

%% file: contents/thread.tex
\subsection{Time efficiency}

With respect to time efficiency, we limit 5 hours for invariant inference for each patch in our dataset. In case invariants of a patch cannot be generated on time, we directly pass the patch to our syntactic classifiers. Meanwhile, assessing the correctness of 139 patches in our evaluation dataset, i.e., Xiong et al. dataset \tool{} took 15.5 hours (i.e., about 7 minutes for each patch). The results show that the assessment time of \tool{} is reasonable but the invariant inference is time-consuming. However, the invariant inference is partially reusable as users can reuse the generated invariants for buggy and patched programs for each patch. Moreover, users can change the time limit for invariant inference if they only have a limited budget. However, even in the worst case, the performance of \tool{} will only drop to the performance of our syntactic classifier, which still outperforms the state-of-the-art baselines. We leave the improvement on time efficiency of invariant inference for future work.

\subsection{Potential Application}
 Although the reliance on ground truth patches limits our applications on pure APR problem settings, \tool{} may be not only useful in patch correctness assessment but also in APR on problem settings where ground truth programs are available. For example, in the context of regression bug fixing~\cite{le2021refixar, tan2015relifix}, a potential ground truth could be the original version before applying a bug-inducing commit. 
 Besides, automated patch correctness assessment with ground truth cannot be directly used in automated program repair, it has been shown to be helpful in the training phase of learning-based program repair, in which the ground truth patches are available~\cite{ye2022neural}.

\subsection{Threats to validity}
\noindent \textbf{External validity}. Threats to external validity correspond to the generalizability of our findings. Our study considers 885 patches generated from 21 popular APR techniques. This may not represent all APR techniques and thus may affect the generalizability of our study. We tried to mitigate this risk by selecting a data set that is commonly used for patch correctness assessment in the APR community~\cite{xiong2018identifying, le2019reliability, wang2020automated, liu2020efficiency}. \modify{Another threat to external validity is that patches in our dataset are only generated for the Defects4J dataset. This may not represent all bugs in real-world projects and thus may affect the generalizability of our findings. Unfortunately, besides Defects4j, there is only one labeled dataset for patch correctness assessment, i.e., QuixBugs.QuixBugs, however, only contains small programs (approximately 35 lines of code on average) that implement basic algorithms such as Depth First Search or Knapsack. These programs differ from our focus in the paper: industrial programs. Meanwhile, obtaining ground truth labels for patches for industrial programs datasets such as Bears and Bugs.jar requires extensive human efforts~\cite{le2019reliability}. Therefore, we would like to leave the evaluation for future work.
}

\vspace{0.2cm}
\noindent \textbf{Internal validity}. Threats to internal validity refer to possible errors in our implementation and experiments. To mitigate this risk, we have carefully re-checked our implementation and experiments.

\vspace{0.2cm}
\noindent \textbf{Construct validity}. Threats to construct validity correspond to the suitability of our evaluation. The main threat in our study is that the correctness of the patches may be subject to subjective bias because they were manually labeled by human annotators, as mentioned in Section ~\ref{sec:auto_patch_assess}. To mitigate this risk, we collected classification results from reliable sources that are widely used in the research community.

%% file: contents/relatedworks.tex
\subsection{Automated Program Repair} Our study investigates patches generated by several popular APR techniques, including \textsc{GenProg}~\cite{le2011genprog}, \textsc{Kali}~\cite{qi2015analysis}, \textsc{Nopol}~\cite{xuan2016nopol}, \textsc{HDRepair}~\cite{le2016history} and \textsc{ACS}~\cite{xiong2017precise}. \textsc{GenProg} and \textsc{Kali} are heuristic-based techniques that construct a search space by using mutation operations and then leverage genetic programming to find the solution. \textsc{Nopol} uses Satisfiability Modulo Theories to synthesize repair for buggy conditional statements. \textsc{HDRepair} mines historical bug-fix patterns to guide the heuristic search. \textsc{ACS} attempts to generate high-quality repairs for buggy conditional statements by using historical fix templates. Beyond these techniques, recently, \textsc{CapGen}~\cite{wen2018context}, \textsc{SimFix}~\cite{jiang2018shaping}, \textsc{FixMiner}~\cite{koyuncu2020fixminer}, and \textsc{TBar}~\cite{liu2019tbar} have been proposed to fix bugs automatically based on frequent fix patterns. Other approaches (e.g., \textsc{SequenceR}~\cite{chen2019sequencer}, \textsc{DLFix}~\cite{li2020dlfix}, \textsc{Coconut}~\cite{lutellier2020coconut}) propose to generate patches by using deep learning models.

\subsection{Overfitting Problem}
Early APR techniques widely leverage test suites, which are often practically weak and incomplete, as an oracle to guarantee patch correctness. This leads to the overfitting problem, in which APR-generated patches pass the validation test suite but are still incorrect~\cite{smith2015cure}. Many APR techniques, e.g., \textsc{GenProg}~\cite{le2011genprog}, \textsc{RSRepair}~\cite{qi2014strength}, \textsc{AE}~\cite{weimer2013leveraging}, and \textsc{Angelix}~\cite{mechtaev2016angelix} have been shown to suffer from the overfitting issue~\cite{qi2015analysis, le2018overfitting}.

\par The overfitting problem has progressively been an important challenge in APR. Monperrus et al. criticized that the conclusiveness of techniques that keep patches and their correctness labels private is questionable~\cite{monperrus2014critical}. Le et al. also suggested making publicly available to the community authors' evaluation on patch correctness~\cite{le2019reliability}. Since then, APR techniques have publicly released their results and labels of APR-generated patches. Authors of APR techniques often assess patch correctness by either using: (1) an independent test suite different from the test suite used for repair to test the generalizability of the generated patches~\cite{le2019reliability}, or (2) manual inspection to compare APR-generated patches with the ground truth~\cite{wen2018context,koyuncu2020fixminer,liu2019tbar,liu2019avatar}. Le et al. show that automated validation via an independent test suite is less effective than manual validation, but there is a potential risk of human bias when using manual validation~\cite{le2019reliability}. Also, manual validation requires repetitive and expensive tasks, which automated validation can complement.

In this work, we use a data set of 885 APR-generated patches for large real-world programs whose correctness labels have been released by recent popular work~\cite{martinez2017automatic, xiong2018identifying, le2019reliability, liu2020efficiency, wang2020automated}. The correctness labels of the patches have been carefully examined by the community, e.g., researchers and independent developers, and thus serve as reliable ground truth labels to assess the effectiveness of APAC techniques that we will discuss next.

\subsection{Automated Patch Correctness Assessment}
To avoid the potential bias of manual patch validation, several techniques have been proposed to predict patch correctness automatically. These techniques can be categorized into different directions: (1) semantic-based APAC and (2) syntactic-based APAC. In this section, we briefly review well-known techniques for each direction.

\subsubsection{Semantic-based APAC}

With respect to semantic-based APAC, the closely related works to our work are \textsc{DiffTGen}~\cite{xin2017identifying} and \rgt{}~\cite{ye2021automated} Similar to our work \tool{}, the techniques identify patch correctness by relying on perfect oracles such as correct programs provided by human developers. To do so, \textsc{DiffTGen} uses \textsc{EvoSuite}, an automated test generation technique to generate an independent test suite from the developer-patched (ground truth) program. \textsc{DiffTGen} considers an APR-generated patch as overfitting if there are any behavioral differences between the APR-patched program and the ground truth program. The fundamental difference between these approaches and \tool{}'s semantic-based classifier is that, instead of generating additional test cases, \tool{} only uses the original test suite and infers program invariants to generalize the desired behaviors of the program under test. This way, \tool{} generates more abstract program specifications in the form of program invariants to effectively guard against unintended behaviors of the programs under test. 

Yu et al.~\cite{yu2019alleviating} also generated additional test cases from the developer-patched program to detect two kinds of overfitting issues: incomplete fixing and regression introduction. However, their approach only works on semantic-based APR techniques while \tool{} can identify overfitting patches generated by all APR approaches. Recently, Yang and Yang explored that the majority of the studied plausible patches (92/96) expose different modifications of runtime behaviors captured by the program invariants, compared to correct patches~\cite{yang2020exploring}. However, this work does not propose any techniques to validate APR-generated patches. Based on the findings of Yang and Yang, Ye et al.~\cite{ye2019comprehensive}, and Wang et al.~\cite{wang2020automated} have also used a simple heuristic based on \textsc{Daikon}'s invariants to identify patch correctness. These heuristics consider a patch as overfitting if it violates any invariants inferred from the developer-patched program. However, developers may add other functions which are unrelated to actual bugs, leading to redundant invariants. Hence, this overfitting behavior is weak and sensitive; that is the reason why they produce many false positives~\cite{ye2019comprehensive}. Meanwhile, \tool{} identifies patch correctness based on carefully designed overfitting behaviors by comparing invariants inferred from both buggy programs and developer-patched programs so that our technique essentially only produces a low false-positive rate, as shown in our evaluation. 

Less relevant to our approach in this work are several techniques attempting to identify patch correctness without knowing perfect oracles. Yang et al.~\cite{yang2017better} proposed \textsc{Opad}, which employs test-suite augmentation based on fuzz testing and uses the crash-free behavior as the oracle to detect overfitting patches. This approach, however, only identifies certain types of overfitting patches such as \textsc{Opad} (as shown in Xiong et al.’s evaluation [43]). Xiong et al.~\cite{xiong2018identifying} proposed \textsc{PATCHSIM} to heuristically identify patch correctness based on the similarity of test case executions. It first uses a test generation tool, i.e., \textsc{Randoop}, to generate new test inputs. It then automatically classifies the generated test cases into passing or failing based on the similarity of execution traces. Finally, it uses an enhanced test suite to determine whether an APR-generated patch is overfitting based on its behaviors on passing and failing test cases. Similar to \textsc{DiffTGen},  \textsc{PATCHSIM} requires the generation of external test cases while \tool{} only uses the original test suite and infers program invariants to generalize the desired behaviors of the program under test. 

\subsubsection{Syntactic-based APAC}

With respect to syntactic-based APAC, the closely related works to our work are \textsc{BERT}+LR proposed by Tian et al.~\cite{tian2020evaluating}. \textsc{BERT}+LR assumes that correct codes differ substantially from incorrect codes and uses code representation techniques to differentiate between them. Specifically, \textsc{BERT}+LR embeds a patched code and a buggy code into numerical vectors using \textsc{BERT}~\cite{devlin2018BERT} and then uses Logistic Regression to estimate the similarity between them. Finally, a patch is considered incorrect/overfitting if the similarity is lower than a certain threshold. However, determining a suitable threshold is challenging because the difference between correct and incorrect codes can vary among programs.
In contrast, our approach considers the similarity of a patched program to its ground truth and buggy program. Our syntactic-based classifier relies on the intuition that a correctly patched code is more similar to the developer-patched code (ground truth) than a buggy code. Thus, the similarity between a patched and ground truth code serves as a “soft threshold” that can be adjusted for different programs. As a result, our approach is more flexible than \textsc{BERT}+LR and achieves better performance, as demonstrated in Section~\ref{sec:eval_finding}. Additionally, our syntactic-based classifier incorporates new syntactic features from CodeBERT~\cite{feng2020codebert}, which has been shown to be more effective than BERT features.

Other works rely on hand-crafted code features to validate the generated patch, including \textsc{Anti-patterns} and \ods{}. Tan et al.~\cite{tan2016anti} propose anti-patterns (i.e., specific static structures) to filter out overfitting patches. Ye et al.~\cite{ye2019automated} leverage 4199 code features extracted from buggy code and generated patches as input to machine learning algorithms (i.e., logistic regression, KNN, and random forest) to rank potentially overfitting patches. However, this work requires manual hand-crafted features that were carefully (manually) engineered, while our approach automatically extracts features via a pre-trained language model.

Different from the aforementioned approaches from both semantic and syntactic-based APAC, our approach leverages both semantic information, i.e., program invariants, and syntactic information, i.e., CodeBERT features, to reason about patch correctness. 

%% file: contents/conclusion.tex
In this paper, we proposed \tool{}, a novel automated patch correctness assessment technique using semantic and syntactic reasoning via program invariants and program syntax. \tool{} first infers program specifications in the form of program invariants, guarding against \emph{correct} and \emph{error} specifications of a program under test. Based on the inferred specifications, \tool{} effectively identifies whether an APR-generated patch is overfitting. In case the above invariant-based specification inference fails to determine an overfitting patch, \tool{} further uses a machine learning model to estimate the probability that the APR-generated patch is overfitting. To do this, \tool{} first uses \textsc{CodeBert}, a well-known pre-trained model of code, to represent the language semantics of program syntax via a vector of numbers and then measures syntactic differences between APR-generated patches and their buggy and correct versions. Based on syntactic differences, \tool{} uses a trained model from labeled patches to estimate the likelihood of an APR-generated patch being overfitting. We compared \tool{} against state-of-the-art automated patch correctness assessment techniques from a popular dataset of 885 APR-generated patches for large real-world projects in \textsc{Defects4J}. Experiment results showed that \tool{} outperforms state-of-the-art baselines.

In future work, we plan to extend \tool{} with other ground truths which are available (e.g., such as the original version of the program before applying a bug-inducing commit). Moreover, the effectiveness of \tool{} demonstrates that program invariants can effectively capture the runtime behaviors of the program. Therefore, another potential direction may be finding a way to take advantage of the program invariants in enhancing automated program repair directly. Finally, we plan to integrate \tool{} as a part of the training process to further improve learning-based program repair, as inspired by RewardRepair~\cite{ye2022neural}.

\vspace{2mm}

\section{Data Availability}
\tool{} is publicly available at \url{https://github.com/thanhlecongg/Invalidator}. All materials including implementation, datasets, and experimental results are also published via \url{https://doi.org/10.5281/zenodo.7699142}